\documentclass[]{jfm}
\usepackage{graphicx}

\usepackage{newtxtext}
\usepackage{newtxmath}
\usepackage{natbib}
\usepackage{hyperref}
\hypersetup{
    colorlinks = true,
    urlcolor   = blue,
    citecolor  = black,
}

\usepackage{epstopdf, epsfig}
\usepackage{caption}
\usepackage{subcaption}
\usepackage{amsmath}
\usepackage{amssymb}
\usepackage{tikz}
\usetikzlibrary{shapes}
\definecolor{blue}{RGB}{0,112,192}
\definecolor{lightblue}{RGB}{0,176,240}
\definecolor{green}{RGB}{0,176,80}
\definecolor{yellow}{RGB}{255,255,0}
\definecolor{orange}{RGB}{255,192,0}
\definecolor{red}{RGB}{255,0,0}
\definecolor{darkred}{RGB}{118,0,0}
\definecolor{purple}{RGB}{208,0,154}

\newcommand{\RomanNumeralCaps}[1]
\linenumbers

\shorttitle{Scaling of granular column collapses on inclined planes}
\shortauthor{T. Man, H. E. Huppert, S. A. Galindo-Torres}

\title{Scaling of granular column collapses on inclined planes}

\author{Teng Man\aff{1},
  Herbert E. Huppert\aff{2}
 \and Sergio A. Galindo-Torres\aff{1}
 \corresp{\email{s.torres@westlake.edu.cn}}
}

\affiliation{\aff{1}Key Laboratory of Coastal Environment and Resources of Zhejiang Province (KLaCER), School of Engineering, Westlake University, 600 Dunyu Rd, Hangzhou, Zhejiang 310024, China
\aff{2}Institute of Theoretical Geophysics, King's College, University of Cambridge, King's Parade, Cambridge CB2 1ST, UK}
\begin{document}

\maketitle

\begin{abstract}
Granular column collapse is a simple but important problem to the granular material community, due to its links to dynamics of natural hazards, such as landslides and pyroclastic flows, and many industrial situations, as well as its potential of analyzing transient and non-local rheology of granular flows. This article proposes a new dimensionless number to describe the run-out behaviour of granular columns on inclined planes based on both previous experimental data and dimensional analysis. With the assistance of the sphero-polyhedral discrete element method (DEM), we simulate inclined granular column collapses with different initial aspect ratios, inter-particle frictions, and initial solid fractions on inclined planes with different inclination angles (2.5$^{\circ}$ - 20.0$^{\circ}$) to verify the proposed dimensional analysis. Detailed analyses are further provided for better understanding of the influence of different initial conditions and boundary conditions, and to help unify the description of the relative run-out distances of systems with different inclination angles. This work determines the similarity and unity between granular column collapses on inclined planes and those on horizontal planes, and helps investigate the transient rheological behaviour of granular flows, which has direct relevance to various natural and engineering systems. 
\end{abstract}


\section{Introduction}
\label{sec:intro}

Understanding the dynamic behaviour of granular flows is crucial for dealing with some natural phenomena \citep{bagnold1954experiments,midi2004dense}, such as debris flows, landslides, and pyroclastic flows \citep{Bougouin2019collapse}, and is also important for solving some engineering issues related to civil engineering \citep{man2021granular,man2023mathematical}, chemical engineering \citep{Ottino2000}, as well as pharmaceutical engineering \citep{Boonkanokwong2021}. While rheological models based on the inertial number, $I$, and the viscous number, $I_v$, successfully describe steady-state behaviours of granular systems \citep{jop2006constitutive,pouliquen2006flow,boyer2011unifying,trulsson2012transition}, most natural and engineering systems are in unsteady state conditions, which may require transient rheological models. The investigation of granular column collapses on either horizontal planes or inclined planes provides us with a simple example of transient granular flows so that both their macroscopic behaviour and local rheological property can be explored accordingly.

\citet{lube2004axisymmetric} and \citet{lajeunesse2005granular} first investigated the dynamics of granular column collapses in a dry condition and on a horizontal plane, and quantified the run-out behaviour of the relationship between the relative run-out distance, $\mathcal{R} = (R_{\infty} - R_i)$, and the initial aspect ratio, $\alpha = H_i/R_i$, where $R_{\infty}$ is the final deposition radius of the axisymmetric granular column, $R_i$ is the initial column radius, and $H_i$ is the initial column height. They concluded that $\mathcal{R}$ approximately scales with $\alpha$ when $\alpha$ is smaller than a threshold $\alpha_c$, and scales with $\alpha^{0.5}$ when $\alpha>\alpha_c$. \citet{staron2005study,staron2007spreading} emphasized the influence of particle properties, such as inter-particle frictional coefficients and coefficients of restitution, with numerical investigations, and found that changing particle properties could influence the energy dissipation process, which lead to different final run-out distances and different collapse kinematics, but did not provide quantitative analyses of these influences. Later, more research has been conducted to study the complexity of granular column collapses with different particle size polydispersities \citep{Cabrera2019granular,Martinez2022segregation}, fluid saturation or immersion condition \citep{rondon2011granular,fern2017granular,Bougouin2019collapse}, different complex particle shapes \citep{Zhang2018influence}, and erodible boundaries \citep{Wu2021collapse}. 

\citet{lube2011granular} first took granular columns onto inclined planes to explore the influence of inclination angles, where they considered 5 different inclination angles ($\theta =$ 4.2$^{\circ}$, 10$^{\circ}$, 15$^{\circ}$, 20$^{\circ}$ and 25$^{\circ}$) and, with dimensional analysis combined with analytical solutions for granular dam-break flows by \citet{mangeney2000analytical}, analyzed run-out behaviours, deposition patterns, and the kinematics of granular columns in different conditions (detailed description of this work will be reviewed in Sec. \ref{sec:prob} since part of this work is directly based on the work of \citet{lube2011granular}). Due to the inclination, granular column collapses on inclined planes exhibit much more complex characteristics. Therefore, it is convenient to use them as a benchmark for verifying certain rheological models or testing different continuum modeling approaches. \citet{Crosta2015} investigated granular column collapses on inclined planes with either erodible or unerodible features with a combined Eulerian-Lagrangian method model. \citet{Chou2023} also studied the erosion and deposition process of granular collapses on an erodible inclined plane, but focused on experimental investigations. \citet{Ionescu2015} used granular column collapses on both horizontal and inclined planes to verify a viscoplastic pressure-dependent rheological model. Similarly, 
\citet{Ikari2016} simulated granular collapses on inclined planes with smooth particle hydrodynamics and the Drucker–Prager yield function, while \citet{Salehizadeh2019} investigated the behaviour of granular column collapses to test their smooth particle hydrodynamics code incoporated with the $\mu(I)$ rheology. \citet{Lee2019underwater} further considered granular column collapses on inclined planes in a subaqueous environment to study the influence of the Darcy number on the behaviour of underwater granular flows. 

However, previous research often lacks physics-based quantitative representation of the influence of frictional properties. Thus, our recent studies introduced an effective initial aspect ratio, 
\begin{equation}
    \begin{split}
        \alpha_{\rm{eff}} = \alpha\sqrt{1/(\mu_w +\beta\mu_p)},
    \end{split}
\end{equation}
where $\mu_w$ is the frictional coefficient between particles and the plane, $\mu_p$ is the inter-particle frictional coefficient, and $\beta = 2$ is a constant, and analyzed the deposition morphology \citep{man2021deposition}, finite-size scaling \citep{man2021finitesize}, as well as the influence of cross-section shapes \citep{man2022influence}, and also introduced a mixture theory to consider the condition when a granular system consists of particles with different frictional properties \citep{man2023bifriction}. Based on our analyses, $\alpha_{\rm{eff}}$ can be seen as a ratio between the inertial effect that drives the granular system forward and frictional influence that dissipates the energy, and the inertial effect can be associated with the energy being transformed from potential to kinetic energy.

In this paper, based on the work of \citet{lube2011granular}, we aim to utilize the previously defined $\alpha_{\textrm{eff}}$ to explore the scaling of granular column collapses on inclined planes with the assistance of the sphero-polyhedral discrete element method. The simulation set-up is similar to that presented in \citet{lube2011granular}. Based on dimensional analysis and simulation results, we are able to relate the relative run-out distance  to a scaling solution, and shed light on the prediction of both dynamic behaviours and deposition patterns of granular column collapses on inclined planes. This paper is organized as follows. Section \ref{sec:prob} provides readers with a detailed description of the problem faced and presents the dimensional analysis for deriving a new dimensionless number with the incorporation of the inclination angle. In Section \ref{sec:DEM}, we describe the simulation set-up and provide the numerical method. We further investigate the influence of inclination angles on flow kinematics and run-out behaviours in Section \ref{sec:flow-n-runout} and the residue height in Section \ref{sec:height}. We will discuss the influence of the initial solid fraction in Section \ref{sec:solidfrac}. Further discussions are introduced in Section \ref{sec:disscussion}, before concluding remarks are made in Section \ref{sec:conclud}.

\section{Problem statement and dimensional analysis}
\label{sec:prob}

\begin{figure}
  \includegraphics[scale=0.11]{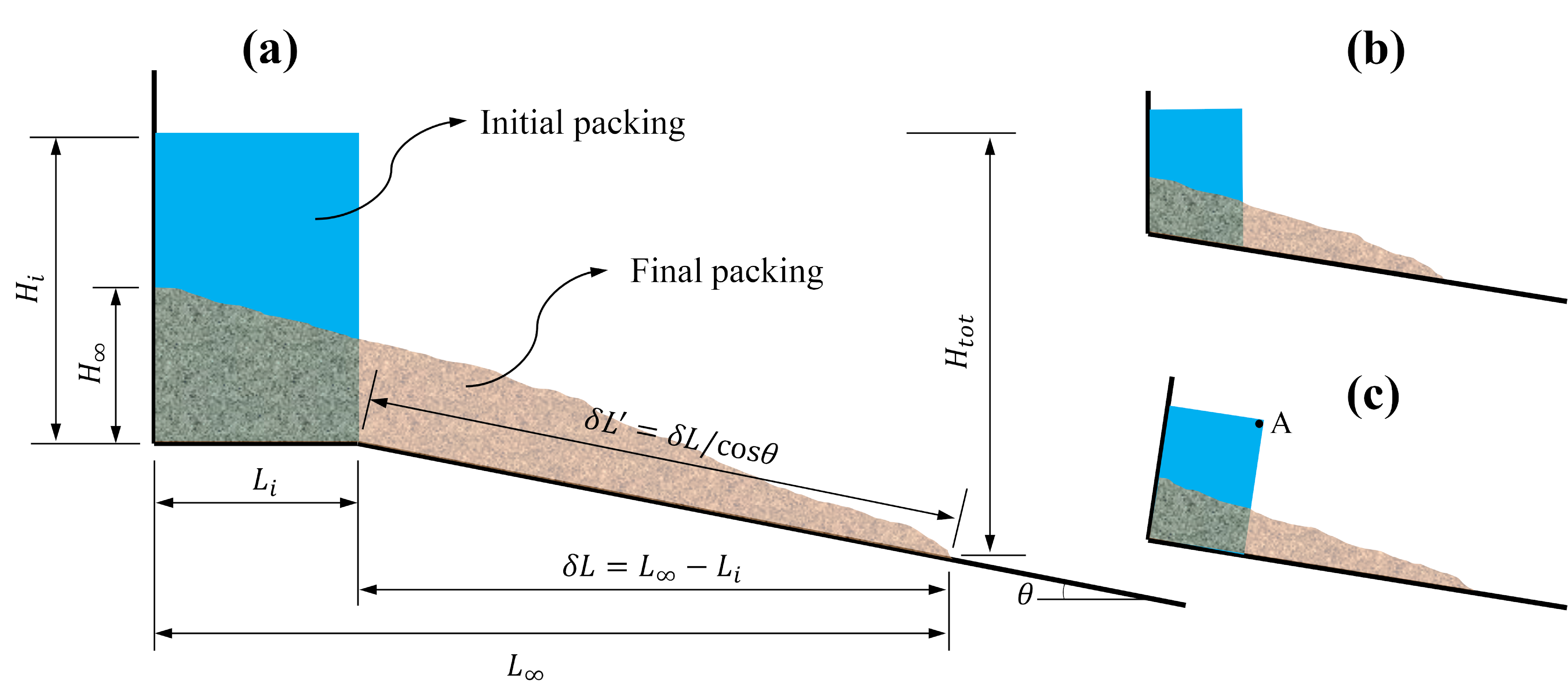}
  \caption{(a) Sketch of the problem set-up, where black lines denote solid boundaries, light blue body represents the initial granular column, and the sand-like body is the final deposition. (b) and (c) show two different type of granular column collapses on inclined planes.}
\label{fig:prob_setup}
\end{figure}

In this work, we aim to investigate granular column collapses on an inclined plane as shown in Figure \ref{fig:prob_setup}(a), which represents a two-dimensional granular column collapse. The initial granular column with height $H_i$ and horizontal length $L_i$ is placed on a horizontal plane. The horizontal plane is connected to an inclined plane with inclination angle $\theta$, so that the initial granular packing (colored blue in Figure \ref{fig:prob_setup}(a)) will collapse onto it once the material is released. After the collapse of the granular column, we can measure the residue deposition height $H_\infty$ and the total horizontal deposition length $L_{\infty}$. Then, we can calculate the horizontal run-out distance $\delta L = L_{\infty} - L_i$ and the inclined run-out distance $\delta L^{\prime} = \delta L/\textrm{cos}\theta$. We are interested in how changing inclination angles and interparticle contact properties influences the behaviour of (i) relative run-out distances, (ii) deposition heights, and (iii) flow kinematics.

We note that our inclined plane is identical to the experimental set-up in \citet{lube2011granular} because this set-up ensures no pre-defined slippery boundary for the column and no free-falling particles exist at the beginning. This set-up is slightly different from the other two options shown in Figures \ref{fig:prob_setup}(b) and \ref{fig:prob_setup}(c), which were often used in previous research \citep{Crosta2015,Chou2023}. In Figure \ref{fig:prob_setup}(b), granular materials are placed vertically on the inclined plane as the initial condition, and the inclined plane beneath it can be regarded as a pre-defined slipping boundary and a possible failure surface, which may influence the run-out results and the deposition pattern. Similarly, in Figure \ref{fig:prob_setup}(c), not only is there a pre-defined slipping boundary for the initial granular packing, but a few particles at the upper right corner (around Point A in Figure \ref{fig:prob_setup}(c)) are initially at a free-fall regime with almost no supporting particles beneath them, which may also influence the collapse phenomenon.

\begin{figure}
  \includegraphics[scale=0.45]{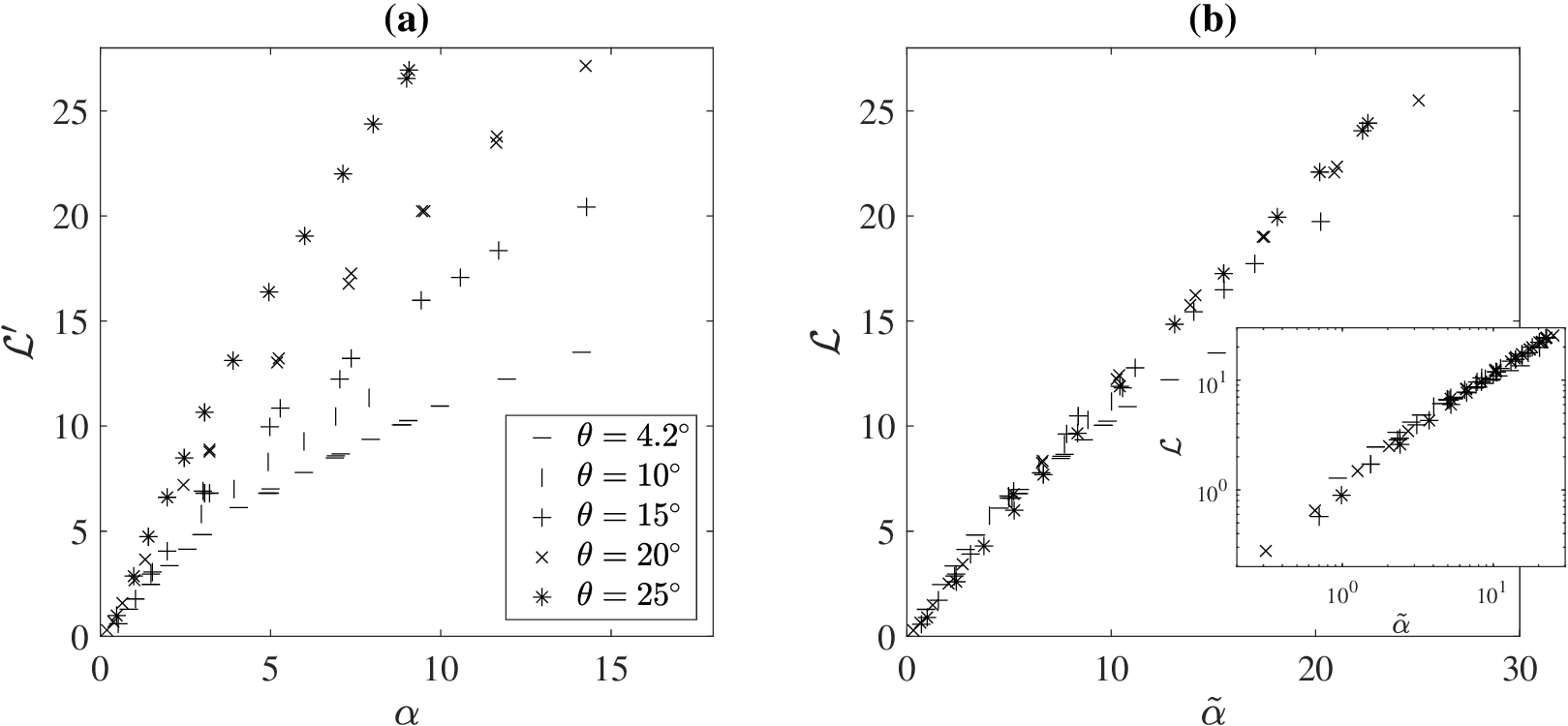}
  \caption{(a) Experimental results extracted from \citet{lube2011granular}. The $y-$axis is the relative run-out distance along the inclination, $\mathcal{L}^{\prime} = \delta L^{\prime}/L_i$ (b) shows the results when we plot the horozontal relative run-out distance, $\mathcal{L} = \delta L/L_i$, against the new dimensionless number, $\Tilde{\alpha}$.}
\label{fig:expres}
\end{figure}

We extract results of run-out distances in \citet{lube2011granular} and plot them in Figure \ref{fig:expres}(a), which shows that changing inclination angles scatters the run-out results. In Figure \ref{fig:expres}(a), the $y$ axis, $\mathcal{L}^{\prime}$, is the relative run-out distance along the inclination, and increasing the inclination angle from 4.2$^{\circ}$ to 25$^{\circ}$ greatly increases the run-out distance. They found that, when $\theta \leq 20^{\circ}$, the run-out distance behaves similar to a system on horizontal planes that the relationship between $\mathcal{L}^{\prime}$ and $\alpha$ scales linearly below a threshold of $\alpha$ and scales with $\alpha^{2/3}$ above that threshold. When the inclination angle is close to the maximum angle of repose of the tested granular material and $\alpha$ is large enough, the power-law will be different. We believe that the increase of the run-out distance is due to two factors: (1) the inclination allows more potential energy to be transformed into kinetic energy, which inevitably increases the run-out distance; (2) the existence of the inclination angle decreases the pressure subjected to the slope from granular materials, which also decreases the resulting frictional effect. These two factors enable more energy for a system to propagate and, meanwhile, reduce the energy dissipation during the column collapse.

In previous works, for granular systems with different frictional properties, we introduced an effective aspect ratio, $\alpha_{\rm{eff}}$, as mentioned in Section \ref{sec:intro}, which denotes the ratio of inertial effects and the frictional dissipation. In this work, we follow this logic but have to adjust both the inertial and the frictional influences. On one hand, since when granular materials flow onto an inclined plane, extra potential energy can be utilized to produce more kinetic energy, which helps the whole system to move forward with a larger run-out distance. One simple hypothesis is that the initial column height should be increased to reflect the change of the available potential energy; thus, we replace $H_i$ with $H_i + \delta L\textrm{tan}\theta$. On the other hand, since the frictional effect is decreased due to the inclination, we introduce a factor of $\cos{\theta}$ to the denominator of the original $\alpha_{\rm{eff}}$, so that a new dimensionless number, which considers both the extra available potential energy and the reduced frictional effect, can be obtained as
\begin{equation}
    \begin{split}
        \Tilde{\alpha}_{\rm{eff}} = \frac{H_i + \delta L\textrm{tan}\theta}{L_i\cdot\textrm{cos}\theta}\sqrt{\frac{1}{\mu_w + \beta\mu_p}}\ \ ,
    \end{split}
\end{equation}
where $\delta L$ is the run-out distance in the horizontal direction. In a word, \citet{lube2011granular} singled out the initial aspect ratio, $\alpha$, and attributed the deviation in $\mathcal{L}^{\prime} - \alpha$ relationship of systems with different inclination angles to the influence of inclinations, but we, in this work, mix the two influences together and investigate the system from a viewpoint of an energy balance. $\Tilde{\alpha}_{\rm{eff}}$ can be named as an inclined effective ratio. We note that \citet{lube2011granular} treated the relative run-out distance along the inclination, $\delta L^{\prime} = \delta L/\textrm{cos}\theta$, as a key result. However, the horizontal and vertical run-out distances are correlated, and it should be the horizontal run-out distance that measures directly the ability of the granular column to transform stored potential energy to kinetic energy. Thus, in this work, we focus on the horizontal run-out distance,$\delta L$, instead of the inclined run-out distance, $\delta L^{\prime}$.

In Figure \ref{fig:expres}(b) and its inset, we plot the relationship between the relative horizontal run-out distance, $\mathcal{L} = \delta L/L_i$, and the new dimensionless number, $\Tilde{\alpha} = (H_i + \delta L\textrm{tan}\theta)/(L_i\textrm{cos}\theta)$. The $x$ axis is $\Tilde{\alpha}$ because the original experiments do not provide the detailed information of particle and boundary frictional coefficients and we simply neglect the part in $\Tilde{\alpha}_{\rm{eff}}$ that constituents frictional coefficients. Most results of inclined granular column collapses with different inclination angles collapse nicely once we plot $\mathcal{L}$ against $\Tilde{\alpha}$, but some deviations appear when $\theta = 4.2$ and $\Tilde{\alpha} > 10$. The inset of Figure \ref{fig:expres}(b) plots the $\mathcal{L} - \Tilde{\alpha}$ relationship in double-logarithmic coordinates, which shows that the $\mathcal{L} - \Tilde{\alpha}$ relationship transforms from one power-law relation to another, as we increase $\Tilde{\alpha}$. This transformation occurs at $\Tilde{\alpha}\approx 3.5$, but the slope change in the log-log plot is not so obvious as that in the $\mathcal{L}-\alpha$ relationship for horizontal granular column collapses.

Figure \ref{fig:expres}(b) shows the applicability and advantage of $\Tilde{\alpha}$ and the possibility of using $\Tilde{\alpha}_{\rm{eff}}$ to quantify granular column collapses on inclined planes with grains of different frictional properties. However, it is difficult to control the particle friction, particle shapes, the boundary friction and the inclination angle in an experiment. Thus, we further investigate this behaviour with numerical methods, so that we can tune both frictional parameters and inclination angles more carefully. 

\section{Discrete element modeling and simulation set-up}
\label{sec:DEM}

\subsection{Sphero-polyhedral discrete element method}

\begin{figure}
  \includegraphics[scale=0.45]{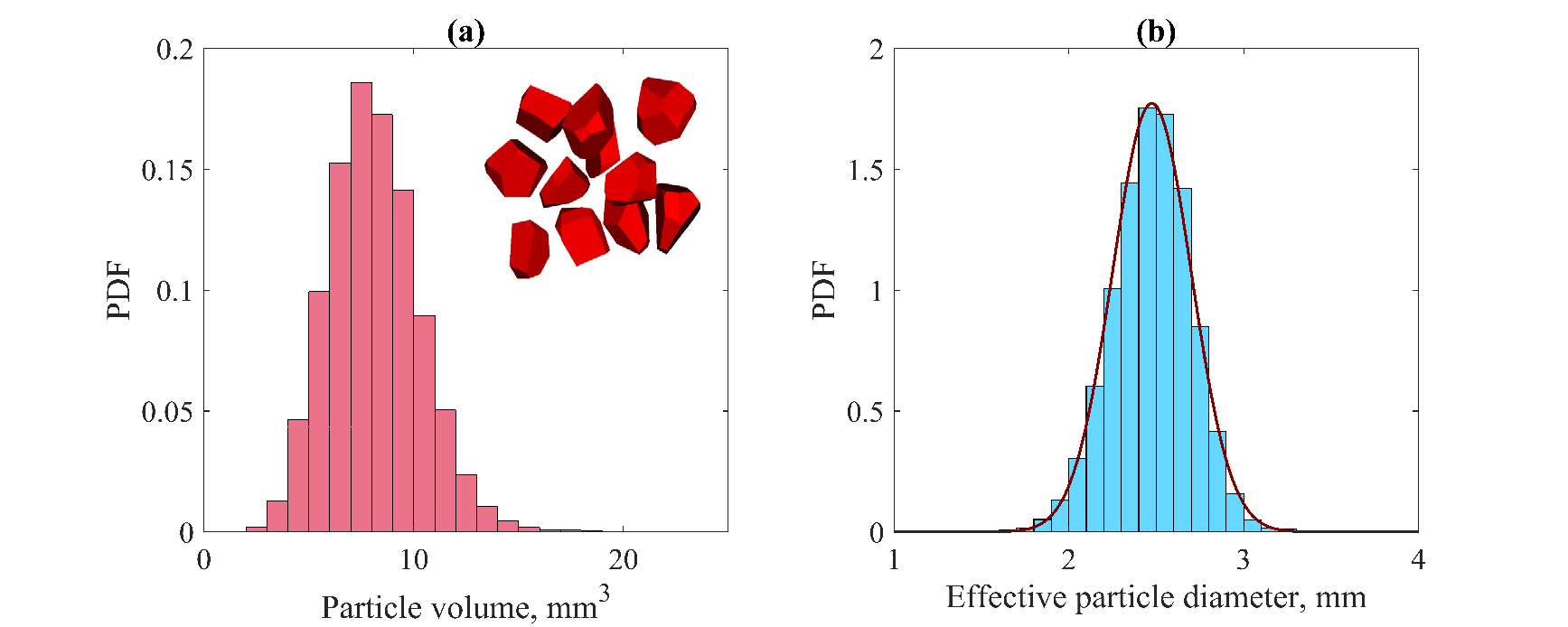}
  \caption{(a) Histogram of particle volumes, $V_{p}$, generated from Voronoi tessellation. The inset shows typical Voronoi-based particles generated from Voronoi tessellation. (b) shows the histogram of the effective particle diameter, $d_{\rm{ep}} = (6V_p/\pi)^{1/3}$.}
\label{fig:psd}
\end{figure}

In this work, we utilize the discrete element method to reflect particle-scale behaviours of granular flows on an inclined plane. A major advantage of the discrete element method is that particle motion is calculated explicitly based on particle contact mechanics and Newton's laws. To use this method, we first need to determine particle shapes and the corresponding contact law. Since we are exploring granular column collapse on inclined planes, we expect that a granular avalanche is initiated and, most importantly, can be stopped naturally. Introducing spherical particles in this system requires us to set up a rolling resistance (both choosing a rolling resistance model and its corresponding parameters), which introduces more parameters that need to be calibrated. Therefore, we naturally choose to generate particles based on the Voronoi tessellation. 

For a simulation, once we identify the initial material domain, a Voronoi tessellation will be performed so that we can obtain a packing of Voronoi-based polyhedrons with initial solid fraction equal to 1. The inset of Figure \ref{fig:psd}(a) shows a few Voronoi-based polyhedra generated from Voronoi tessellation. Figure \ref{fig:psd}(a) shows the histogram of volumes of approximately 36 thousands particles generated from Voronoi tessellation within a $3\times3\times40$ cm$^3$ domain. We see that most particle volumes are in the range between 3 mm$^3$ and 15 mm$^3$ with mean volume of approximately 8.28 mm$^3$ and median volume of 8 mm$^3$. The standard deviation of generated particle volumes is 7.65 mm$^3$.

It is difficult to conclude a possible size distribution function for particle volumes, but the effective particle diameter, as shown in Figure \ref{fig:psd}(b), clearly follows a normal distribution, as shown by the solid curve in Figure \ref{fig:psd}(b). An effective particle diameter, $d_{\rm{ep}}$, is calculated based on regarding each polyhedron as a sphere with the same volume, so that $d_{\rm{ep}} = (6V_p/\pi)^{1/3}$, where $V_p$ is the particle volume. Most $d_{\rm ep}$'s fall between 2 mm and 3 mm with the mean value equal to 2.475 mm and the standard deviation of approximately 0.225 mm. The randomness of both particle shapes and particle sizes ensures that no granular crystallization will be formed during the granular column collapse.

We calculate the contact between Voronoi-based particles based on the sphero-polyhedral method, where each polyhedron is eroded and dilated by a spherical element to obtain a particle with similar shape as the original polyhedron, but with rounded edges and corners, as discussed in \citet{galindo2013coupled} and \citet{man2023bifriction}. The contact between two Voronoi-based particles can be then calculated based on the overlap $\delta_n$, relative tangential displacement vector $\bf{\Xi}$ and relative normal velocity vector $\bf{v}_n$ between contacting spherical elements. The normal and tangential forces between two contacting Voronoi-based particles are calculated as
\begin{subequations}
    \begin{align}
        \bf F_n = \it -K_n\delta_n\bf\hat{n} -\it m_e\gamma_{n}\bf{v}_n,\\
        \bf{F}_t = -\textrm{min}\left(\it |K_t\bf{\Xi}|,\ \it\mu_p |\bf{F}_n|\right)\hat{t},
    \end{align}
\end{subequations}
where $K_n$ and $K_t$ are normal and tangential stiffness of particles, $m_e = 0.5(1/m_1 + 1/m_2)^{-1}$ is the reduced mass, $m_1$ and $m_2$ are masses of contacting particles, respectively, $\mu_p$ is the frictional coefficient of particle interactions, $\bf\hat{n}$ and $\bf\hat{t}$ are unit vectors of normal and tangential direction, and $\gamma_n$ is the normal energy dissipation constant, which depends on the coefficient of restitution $e$ as \citep{alonso2013experimental,galindo2018micromechanics},
\begin{equation} \label{eq:restitution}
        e = \textrm{exp}\left(-\frac{\gamma_n}{2} \frac{\pi}{\sqrt{\frac{K_n}{m_e} - (\frac{\gamma_n}{2})^2}}\right)\ .
\end{equation}
The motion of particles is then calculated by step-wise resolution of Newton's second law with the normal and contact forces using the velocity-Verlet method \citep{Scherer2017}. The discrete element method was incorporated in an open-source computing package, MechSys, developed and maintained by one of authors of this work \citep{galindo2013coupled}, and was validated by various peer-reviewed articles \citep{galindo2010molecular,man2021deposition, man2022influence}. 

\subsection{Simulation set-up}

\begin{figure}
  \includegraphics[scale=0.42]{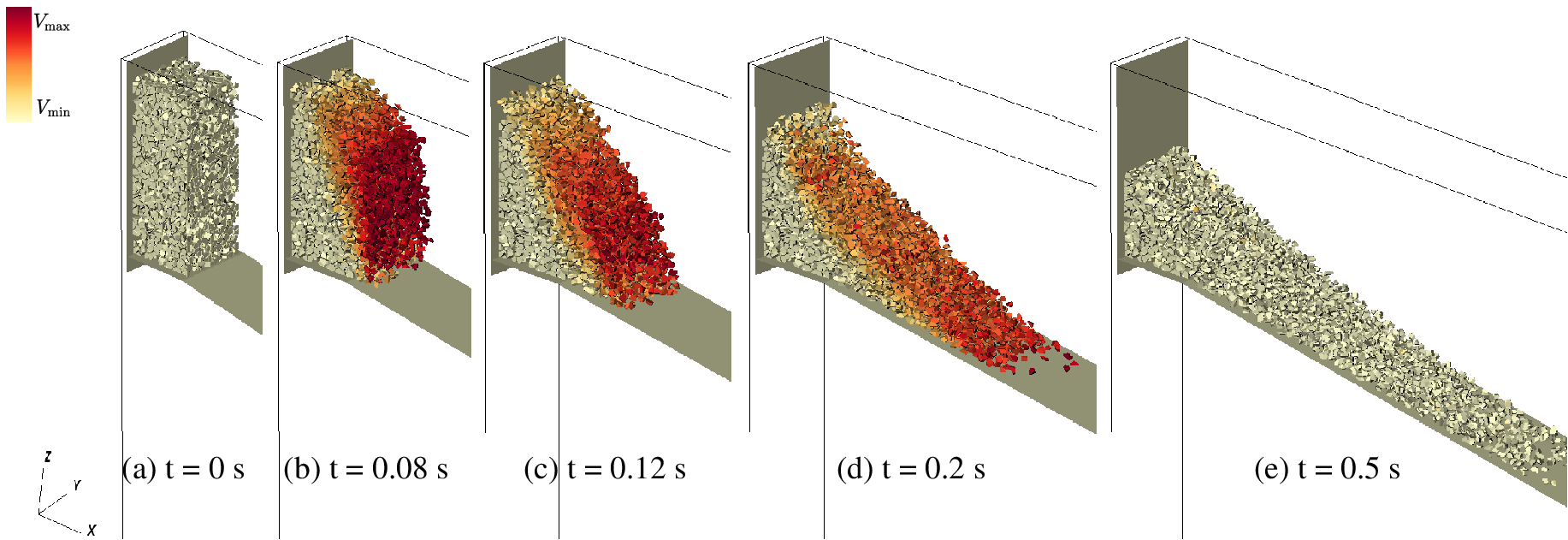}
  \caption{A discrete element simulation of granular column collapses onto an inclined plane of $\theta = 10^{\circ}$. Snapshots are taken at (a) $t = 0$ s, (b) $t = 0.08$ s, (c) $t = 0.12$ s, (d) $t = 0.2$ s, and (e) $t = 0.5$ s. The $x$ axis is toward the horizontal direction, and the $z$ axis is toward the vertical direction. Different colours represent different velocity magnitudes of particles. The colour bar in the figure shows the range of colour that corresponds to the velocity magnitude varying from 0 to its maximum.}
\label{fig:simuset}
\end{figure}

The simulation set-up is similar to that in the experiment presented in \citet{lube2011granular}, but we can explicitly control the frictional properties of both boundaries and particles and set up periodic boundary conditions. We show the simulation set up in Figure \ref{fig:simuset}. The $x$ axis is in the horizontal direction, the $z$ direction is in the vertical direction, and the $y$ axis is pointing into the $x-z$ plane. The simulation has three boundary plates: (1) a vertical plate with the frictional coefficient of $\mu_{bv} = 0$ so that the collapsing granular materials will not face resistance from the vertical wall, (2) a horizontal plane, on which we place the granular packing at the initial state, and (3) an inclined plane, which the granular column will collapse onto once we release particles. The length of the horizontal plate along the $x$ axis is the initial horizontal length, $L_i$, of the granular column. The boundaries vertical to the $y$ axis are periodic boundaries. The distance between two periodic boundaries is the width, $W_i$, of the two dimensional column collapse; and we set $W_i = 3$ cm. 

At the initial condition shown in Figure \ref{fig:simuset}(a), we identify the initial domain of the granular column within $L_i \times W_i \times H_i$ and perform the Voronoi tessellation to form a Voronoi granular packing with solid fraction 1. To make sure that granular materials are loosely packed at the initial state, we choose to reduce the initial solid fraction to $\phi_{\rm init} = 0.6$ by randomly removing 20\% of grains from the Voronoi tessellation. The removal of grains has almost no influence on the mean value and the standard deviation of the effective particle diameter of the granular system. Then, we release the packing and let it flow onto the incline plane. We set the boundary frictional coefficients on the horizontal plate and on the incline plane at the same value, which is $\mu_w = 0.4$. Figures \ref{fig:simuset}(b-e) show the initiation, propagation and termination of the collapse of a granular column with $H_i = 10$ cm. During the collapse process, we record the translational and angular velocities, positions, translational and rotational kinetic energies and particle interactions of the system. We also measure the front velocity during the collapse and determine the terminal time, $T_f$, based on the magnitude of the front velocity.

After the flow termination, we measure the final horizontal length, $L_{\infty}$, and the deposition height, $H_{\infty}$, of the granular pile, to obtain parameters shown in Figure \ref{fig:prob_setup}(a). In this work, to quantify the propagation capacity of granular column collapses, we focus on the horizontal relative run-out distance, $\delta L$, instead of the run-out distance along the inclination, $\delta L^{\prime}$. We note that the way we generate the initial packing leads to a much more stable initial state than loosely packed sand. The initial Voronoi-based packing with $\phi_{\rm init}$ is similar to a fissured porous rock. The face-to-face interactions naturally dominate inter-particle contacts at the initial state, which results in a more stable status for the granular packing. This indicates that the scaling results of simulated granular column collapses may be different from experimental results obtained by \citet{lube2011granular}, but the underlying physics should be similar. In this work, to investigate the scaling of the run-out behaviour and kinematics of granular column collapses on inclined planes, we set up three different inter-particle frictional coefficients, which are 0.2, 0.4 and 0.6, vary the initial height from 1 cm to 50 cm, so that the initial aspect ratio, $\alpha = H_i/L_i$, varies from 0.33 to 16.67, and change the inclination angle, $\theta$, from $2.5^{\circ}$ to $20^{\circ}$.

\section{Run-out behaviour and flow kinematics}
\label{sec:flow-n-runout}

\begin{figure}
  \includegraphics[scale=0.45]{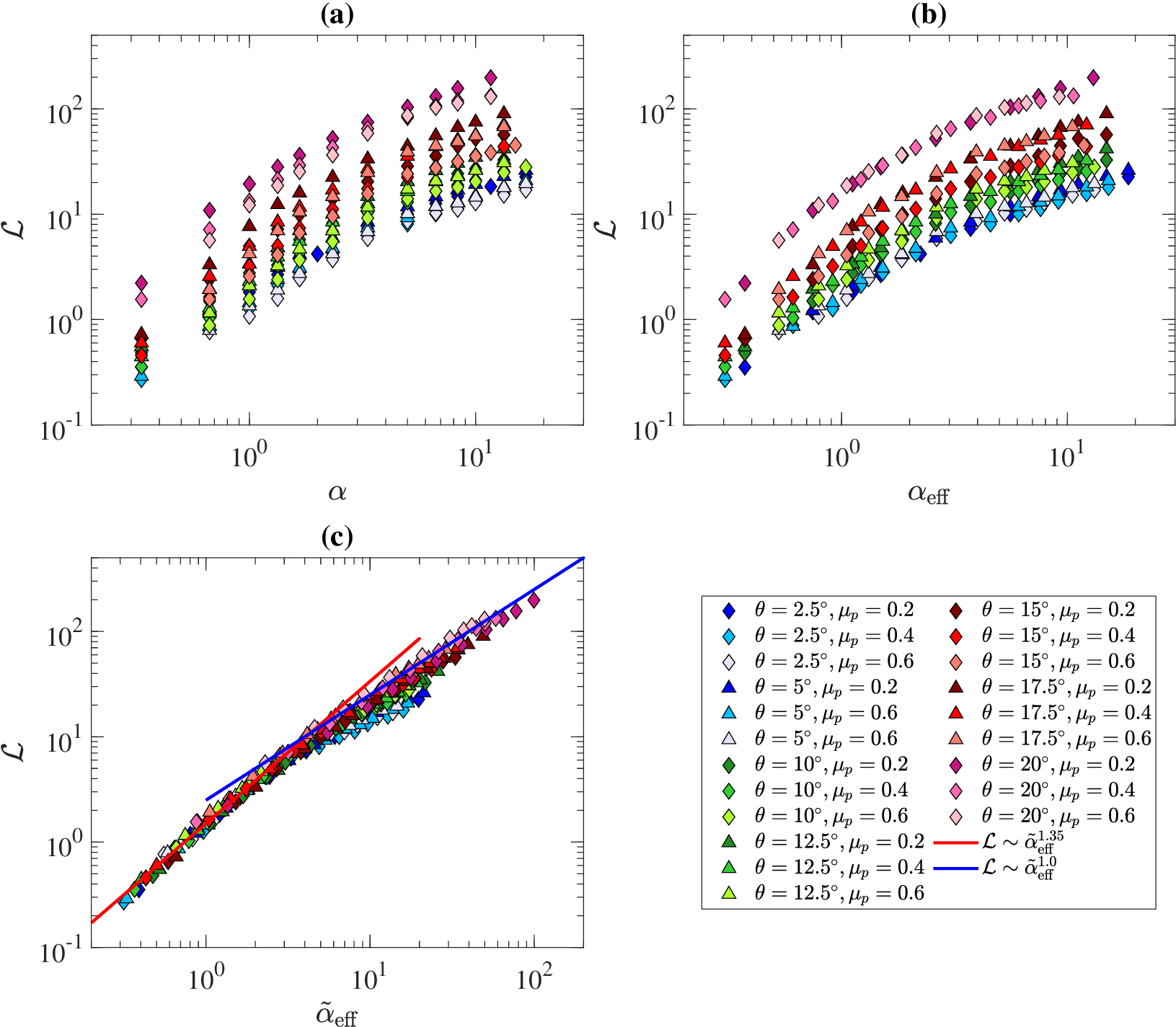}
  \caption{Relative horizontal run-out distance of systems with $\phi_{\rm init} = 0.6$ plotted against (a) initial aspect ratios, $\alpha$, (b) effective aspect ratios, $\alpha_{\rm{eff}}$, and (c) inclined effective aspect ratio, $\Tilde{\alpha}_{\rm{eff}}$, for 21 different sets of simulations. The red curve represents the fitting relationship of $\mathcal{L}\sim\Tilde{\alpha}_{\rm eff}^{1.35}$ and the blue curve denotes the fitting of $\mathcal{L}\sim\Tilde{\alpha}_{\rm eff}$}
\label{fig:runout0.6}
\end{figure}

\subsection{Horizontal run-out distance}

The final run-out distance is a major property for granular column collapses since it exhibits the ability of a granular system to transform potential energy to kinetic energy and links to propagation capacity and damage level of geophysical flows, such as landslides and pyroclastic flows, in natural systems. \citet{lube2011granular} examined the run-out behaviour of granular column collapses on inclined planes and treated the run-out distance on the inclination as a key parameter. However, in this work, we focus on the horizontal run-out behaviour and treat the vertical run-out as a result of both horizontal run-out distance and the inclination angle, and regard the horizontal run-out distance as a direct measurement quantifying the transformation from potential to kinetic energy.

We plot the relative horizontal run-out distance, $\mathcal{L} = (L_{\infty} - L_i)/L_i = \delta L/L_i$, against the initial aspect ratio, $\alpha = H_i/L_i$, of systems with $\phi_{\rm init} = 0.6$ in Figure \ref{fig:runout0.6}(a). For each set of simulations with the same inclination angle $\theta$, we have three different inter-particle frictional properties, i.e., $\mu_p =$ 0.2, 0.4 and 0.6. We can see from Figure \ref{fig:runout0.6}(a) that increasing the inter-particle frictional coefficient from 0.2 to 0.6 helps decrease the run-out distance, and changing frictional properties scatters corresponding data points. Increasing the inclination angle greatly increases the run-out distance. For instance, for a systems with $\alpha \approx 1.0$ and $\mu_p = 0.4$, $\mathcal{L}$ is less than 1 when $\theta = 2.5^{\circ}$, but $\mathcal{L}$ is already larger than 10 when $\theta = 20^{\circ}$. This indicates that increasing $\theta$ by only a factor of 8 results in a run-out distance more than 10 times longer, which implies that the relationship between the initial aspect ratio and the relative run-out distance is nonlinear.

Utilizing the effective aspect ratio $\alpha_{\rm eff}$ that we proposed previously, we are able to collapse the simulation results of systems with the same $\theta$ onto one curve as shown in Figure \ref{fig:runout0.6}(b). The ability of $\alpha_{\rm eff}$ to quantify horizontal granular column collapses has been verified in many previous works \citep{man2021deposition,man2021finitesize,man2022influence,man2023bifriction}, and it still works for a system with an inclined run-out. However, $\alpha_{\rm eff}$ fails to combine all the data into a master curve since the influence of $\theta$ is missing in the definition of it, but most importantly, changing the $x$ axis from $\alpha$ to $\alpha_{\rm eff}$ imposes no effect to the nature that larger inclination angles lead to longer run-out distances. Based on the analysis in Section \ref{sec:prob}, we hypothesize that the inclined effective ratio, $\Tilde{\alpha}_{\rm eff}$, which includes both frictional properties and the inclination information, could help quantify and unify the relationship between run-out distances and initial geometries.

\begin{figure}
  \centering
  \includegraphics[scale=0.45]{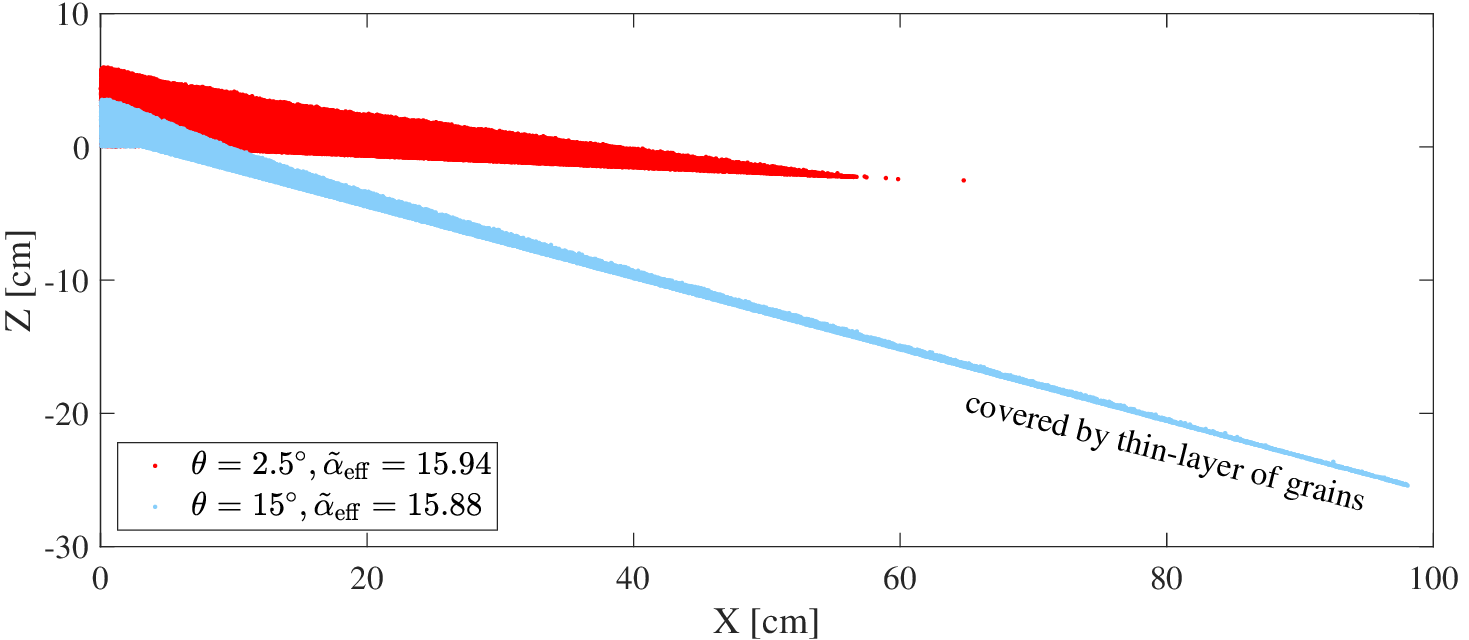}
  \caption{Deposition pattern for granular column collapses with $\theta = 2.5^{\circ}, H_i = 50$ cm, $\Tilde{\alpha}_{\rm eff} = 15.94$ (as red dots) and $\theta = 15^{\circ}, H_i = 25$ cm, $\Tilde{\alpha}_{\rm eff} = 15.88$ (as light blue dots).}
\label{fig:deposit}
\end{figure}

Figure \ref{fig:runout0.6}(c) plots the relationship between $\mathcal{L}$ and $\Tilde{\alpha}_{\rm eff}$ in a log-log coordinate system for simulation results obtained from granular column collapses with $\theta = 2.5^{\circ} - 20^{\circ}$. As expected, changing the $x$ axis to $\Tilde{\alpha}_{\rm eff}$ helps tremendously, in that almost all the data points fall onto a master curve, except for systems with $\Tilde{\alpha}_{\rm eff}$ and $\theta \leq 5^{\circ}$. The $\mathcal{L} - \Tilde{\alpha}_{\rm eff}$ relationship consists of two parts. When $\Tilde{\alpha}_{\rm eff} \lessapprox 4$, $\mathcal{L}$ approximately scales with $\Tilde{\alpha}_{\rm eff}^{1.35}$ as shown by the red line in Figure \ref{fig:runout0.6}(b), but when $\Tilde{\alpha}_{\rm eff} \gtrapprox 4$, $\mathcal{L}$ approximately scales proportionally to $\Tilde{\alpha}_{\rm eff}$, as shown by the blue line in this figure. The parameters of the power-law scaling are different from those in the $\mathcal{L} - \alpha_{\rm eff}$ relationship reported by \citet{man2021deposition} since $\Tilde{\alpha}_{\rm eff}$ contains information of the final run-out distance $L_{\infty}$ inside its definition. When $\Tilde{\alpha}_{\rm eff}$ and $\theta \leq 5^{\circ}$, the simulation results slightly deviate from other data points, and the master curve would over-predict the run-out distance. In one of our previous works \citep{man2021deposition}, we classified granular column collapses into three different regimes: quasi-static, inertial, and fluid-like. One key characteristic of a granular collapse within the fluid-like regime is that the inertial effect becomes so large that the memory of the initial packing, i.e., the initial contact structure and the initial geometry, will be lost, which results in the behaviour that particles initially at the bottom of the packing flow to the very front of the final deposition pile. The data deviation of granular columns with $\Tilde{\alpha}_{\rm eff}$ and $\theta \leq 5^{\circ}$ implies that the fluid-like regimes for systems with different inclination angles might be different from each other.

We choose two cases to investigate and plot the deposition pattern on the $X-Z$ plane in Figure \ref{fig:deposit}. The red dots (or the red region) represent the deposition pattern of a granular column collapse with $\theta = 2.5^{\circ}, \mu_p = 0.4, H_i = 50$ cm, $\Tilde{\alpha}_{\rm eff} = 15.94$, and the blue dots show the pattern of a collapse with $\theta = 15^{\circ}, \mu_p = 0.4, H_i = 25$ cm, $\Tilde{\alpha}_{\rm eff} = 15.94$. Two systems have similar $\Tilde{\alpha}_{\rm eff}$ and both reached the fluid-like regime, as we have argued in \citet{man2021deposition}, but have different inclination angles. The red dots show that, after the granular column collapse, the granular pile is similar to a horizontal granular column collapse with triangular-like deposition pattern. However, for a granular column collapse with a larger inclination angle, the deposition structure becomes different. The blue dots show that a large part of the deposition is covered by only one or two layers of particles, which indicates a larger relative run-out distance than systems with $\theta = 2.5^{\circ}$ and 5$^{\circ}$. We believe that it is the ability to generate a large area of one-layer particle cover that results in the deviation in the $\mathcal{L} - \Tilde{\alpha}_{\rm eff}$ plot. In the system represented by red dots, the thin-layered region is small compared to the length of the granular pile. We can find one particle that reaches $X = 80$ cm and a few particles present between $X = 60$ cm and $X = 80$ cm, but those particles are all detached from the main pile and cannot be regarded as a thin-layered area.

\subsection{Kinetic energy}

\begin{figure}
  \centering
  \includegraphics[scale=0.45]{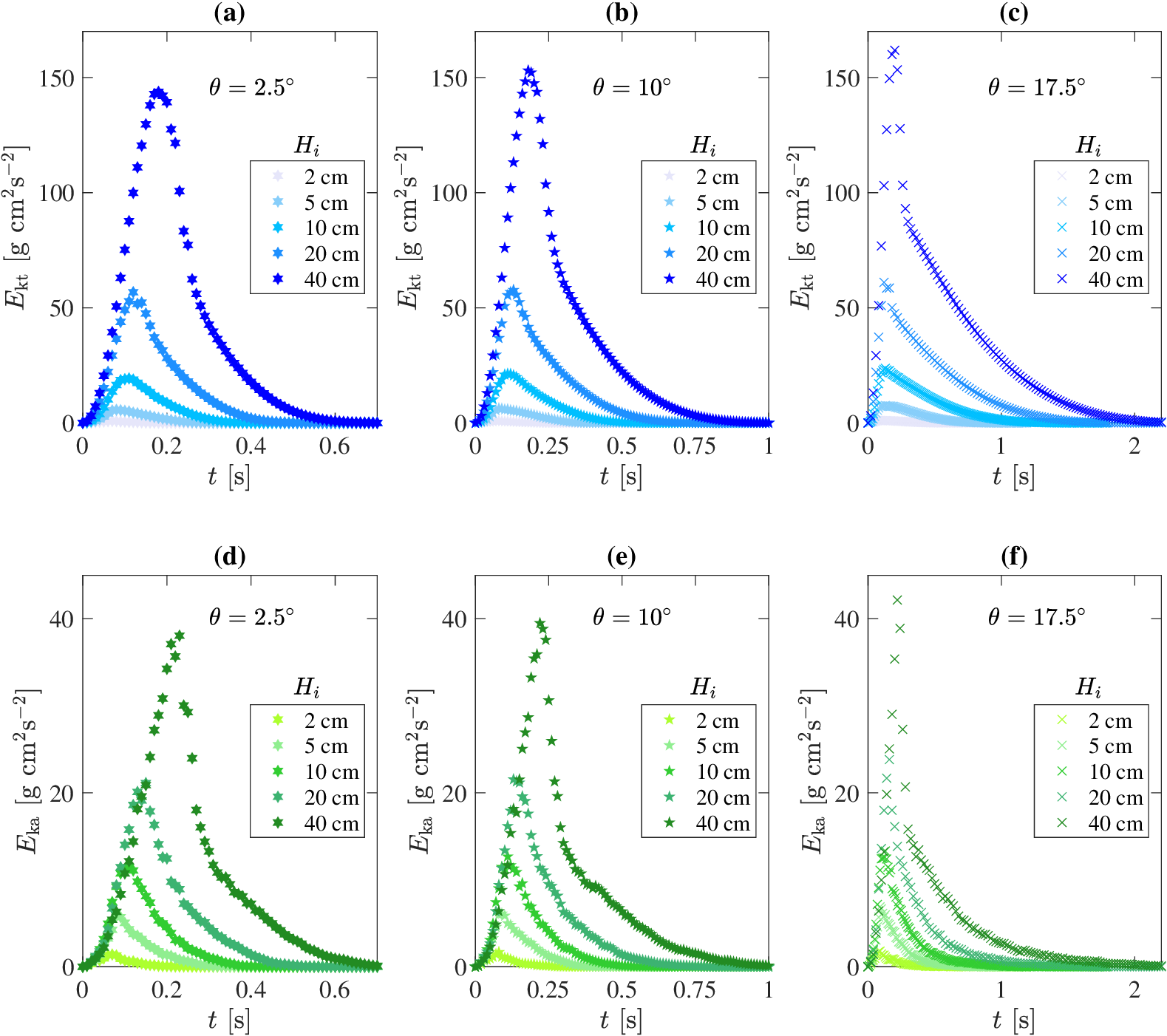}
  \caption{(a - c) Time evolution of the translational kinetic energy per particle, $E_{\rm kt}$, for systems with $\theta = 2.5^{\circ}$, $\theta = 10^{\circ}$, and $\theta = 17.5^{\circ}$, respectively. We only choose columns with five different initial height ($H_i = $2 cm, 5 cm, 10 cm, 20 cm and 40 cm) and $\mu_p = 0.4$ to plot. (d - f) Time evolution of the rotational kinetic energy per particle, $E_{\rm ka}$, for systems with $\theta = 2.5^{\circ}$, $\theta = 10^{\circ}$, and $\theta = 17.5^{\circ}$, respectively. We also choose columns with five different initial height ($H_i = $2 cm, 5 cm, 10 cm, 20 cm and 40 cm) to plot.}
\label{fig:EkTime}
\end{figure}

We have found that increasing the inclination angle leads to considerable increase in the run-out distance and results in a long run-out tail with a thin layer of particles. In this section, we further explore this behaviour from the viewpoint of the energy transformation. At each time, we record the translational and angular velocity vector of each particle, $\bf{v}_{\it p}$ and $\bf \Omega_{\it p}$, and calculate its corresponding translational and rotational kinetic energy based on its mass, $m_p$, and its inertia matrix, $\textbf{I}_p$. Then, we quantify the kinetic energy per particle of this system using the following equations,
\begin{subequations}
    \begin{align}
        E_{kt} = \frac{1}{N_p}\sum_{p\in N_p}^{N_p}\left(\frac{1}{2} m_p\textbf{v}_p^2 \right) \ , \\
        E_{ka} = \frac{1}{N_p}\sum_{p\in N_p}^{N_p}\left(\frac{1}{2} \textbf{I}_{p}\bf{\Omega}_{\it p}\cdot\bf{\Omega}_{\it p} \right) \ , 
    \end{align}
\end{subequations}
where $E_{kt}$ and $E_{ka}$ are translational and rotational kinetic energy per particle, respectively, and $N_p$ is the number of particles in the granular column collapse system. In Figure \ref{fig:EkTime}(a-c), we plot the time evolution of $E_kt$ for granular columns with $\theta = 2.5^{\circ}$, $10^{\circ}$ and $17.5^{\circ}$. We choose systems with five different initial heights to plot. The time evolution of the particle kinetic energy resembles the granular column collapses on a horizontal plane. At the beginning of a collapse, $E_{kt}$ increases nonlinearly with respect to $t$, after which $E_{kt}$ increases rapidly to its peak. Afterwards, the kinetic energy per particle starts to decline and exhibits an exponential decay. As we increase the initial height from 2 cm to 4 cm, the maximum translational kinetic energy increases, as does the duration of the non-zero $E_{kt}$ period. If we compare systems with different inclination angles, we can see that increasing the inclination angle does not result in much increase in the maximum translational kinetic energy, $E_{\rm kt,max}$. For instance, when $H_i = 40$ cm and $\theta = 2.5^{\circ}$, as shown in Figure \ref{fig:EkTime}(a), the maximum translational kinetic energy $E_{\rm{kt,max}} \approx 145$ g$\cdot$cm$^2$s$^{-2}$. As we increase the inclination angle to 10$^{\circ}$, $E_{\rm{kt,max}}$ only increases to approximately 155 g$\cdot$cm$^2$s$^{-2}$. Further, increasing $\theta$ to 17.5$^{\circ}$ only manages to increase $E_{\rm{kt,max}}$ to $\approx$ 165 g$\cdot$cm$^2$s$^{-2}$.

\begin{figure}
  \centering
  \includegraphics[scale=0.45]{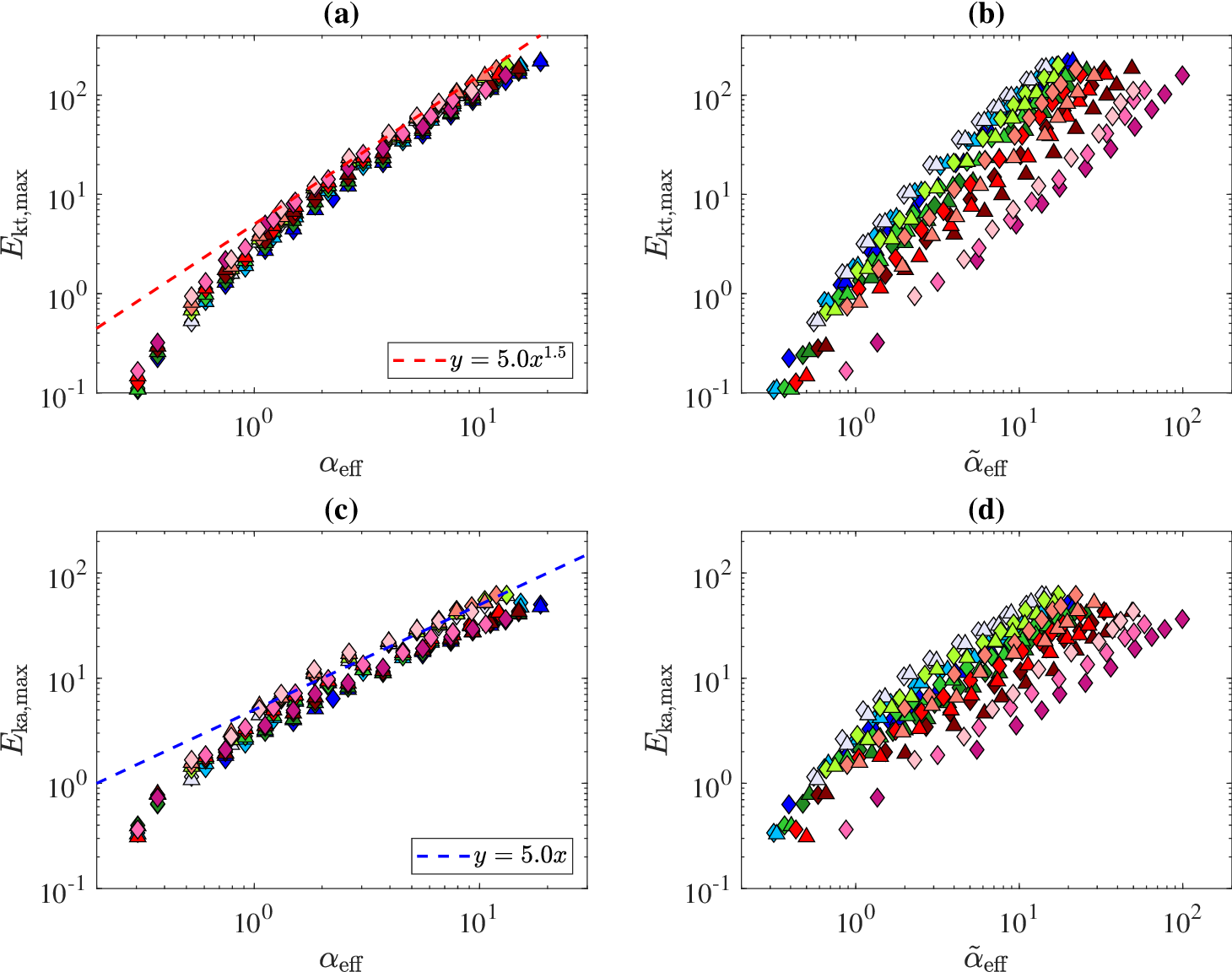}
  \caption{(a) The relationship between the maximum translational kinetic energy per particle in each simulation, $E_{\rm{kt,max}}$, and $\alpha_{\rm eff}$. (b) The relationship between $E_{\rm{kt,max}}$ and $\Tilde{\alpha}_{\rm eff}$. (c) The relationship between the maximum rotational kinetic energy per particle in each simulation, $E_{\rm{ka,max}}$, and $\alpha_{\rm eff}$. (d) The relationship between $E_{\rm{ka,max}}$ and $\Tilde{\alpha}_{\rm eff}$. Markers in this figure are the same with those in Figure \ref{fig:runout0.6}.}
\label{fig:EkMax}
\end{figure}

Similar behaviour happens when we analyze the rotational kinetic energy and its maximum for systems with different initial heights and inclination angles, where increasing $\theta$ from 2.5$^{\circ}$ to 17.5$^{\circ}$ only leads to an increase of $E_{\rm{ka,max}}$ approximately from 38 g$\cdot$cm$^2$s$^{-2}$ to 42 g$\cdot$cm$^2$s$^{-2}$. This phenomenon may result from the fact that, during a granular column collapse, most of the granular system halts quickly after the release of materials and it is mainly the front part that is propagating, which often results in a long thin-layer of particles, as shown in Figure \ref{fig:deposit}. Additionally, this behaviour implies that the major influence of the angle inclination is to extend the collapse duration rather than to increase $E_{\rm{kt,max}}$ or $E_{\rm{ka,max}}$. Nevertheless, we extract the data of $E_{\rm{kt,max}}$ and $E_{\rm{ka,max}}$ for each simulation, and plot them against $\alpha_{\rm eff}$ and $\Tilde{\alpha}_{\rm eff}$ in Figure \ref{fig:EkMax}.

Figure \ref{fig:EkMax}(a) shows on a logarithmic coordinate system the relationship between the maximum particle kinetic energy, $E_{\rm kt,max}$, and the effective aspect ratio, $\alpha_{\rm eff}$. The $E_{\rm kt,max} - \alpha_{\rm eff}$ relation seems to collapse well, which confirms that changing the inclination angle has almost no influence on $E_{\rm kt,max}$. We note that, in Figure \ref{fig:EkMax}(a), the $x$ axis is $\alpha_{\rm eff} = \alpha\sqrt{1/(\mu_w + \beta\mu_p)}$, which bears no $\theta-$related influences. As we increase $\alpha_{\rm eff}$, $E_{\rm kt,max}$ gradually converges to  a power-law relationship that scales with $\alpha_{\rm eff}^{1.5}$. The convergence point $\alpha_{\rm eff}\approx 2$ coincides with turning points in the $\mathcal{L} - \alpha_{\rm eff}$ relationships shown in Figure \ref{fig:runout0.6}(b), where slopes change at $\alpha_{\rm eff}\approx 2$ for almost all sets of simulations with different inclination angles. Figure \ref{fig:EkMax}(b) shows the failure of $\Tilde{\alpha}_{\rm eff}$ in terms of collapsing data of $E_{\rm kt,max}$. It shows that, when $\Tilde{\alpha}_{\rm eff}\gtrapprox 1$,  $E_{\rm kt,max}$ has a power-law scaling with respect to $\Tilde{\alpha}_{\rm eff}$. However, when $\Tilde{\alpha}_{\rm eff}\lessapprox 1$, the slope of the $E_{\rm kt,max}$ - $\Tilde{\alpha}_{\rm eff}$ relationship on the log-log coordinate becomes larger. This behaviour is similar for the maximum rotational kinetic energy, $E_{\rm ka,max}$. 

Figures \ref{fig:EkMax}(c) and (d) show the success of $\alpha_{\rm eff}$ and the failure of $\Tilde{\alpha}_{\rm eff}$ to quantify $E_{\rm ka,max}$. Despite the scatter of the $E_{\rm ka,max}$ - $\Tilde{\alpha}_{\rm eff}$ relationship, the $E_{\rm ka,max}$ - ${\alpha}_{\rm eff}$ relationship collapses well onto a master curve, where the $E_{\rm ka,max}$ gradually converges to a linear curve as we increase ${\alpha}_{\rm eff}$. We can also find out which kinetic energy is dominating the collapsing process from Figure \ref{fig:EkMax}(a) and (c). When $\alpha_{\rm eff} \approx 0.3$, $E_{\rm kt,max}$ is between 0.1 and 0.2, while $E_{\rm ka,max}$ is approximately 0.4. This indicates that, when $\alpha_{\rm eff}$ is small, most of the potential energy will be transformed into rotational kinetic energy. This corresponds to the quasi-static collapse illustrated in \citet{man2021deposition}, where the granular column slumps like a viscous solid and particles often roll down the granular slope, since the effective shear rate and its corresponding stress are not large enough to overcome the frictional interaction between contacting pairs. When $\alpha_{\rm eff} \approx 2$, $E_{\rm kt,max}$ and $E_{\rm ka,max}$ become almost equal, after which the translational kinetic energy dominates in the collapse process.

\begin{figure}
  \centering
  \includegraphics[scale=0.45]{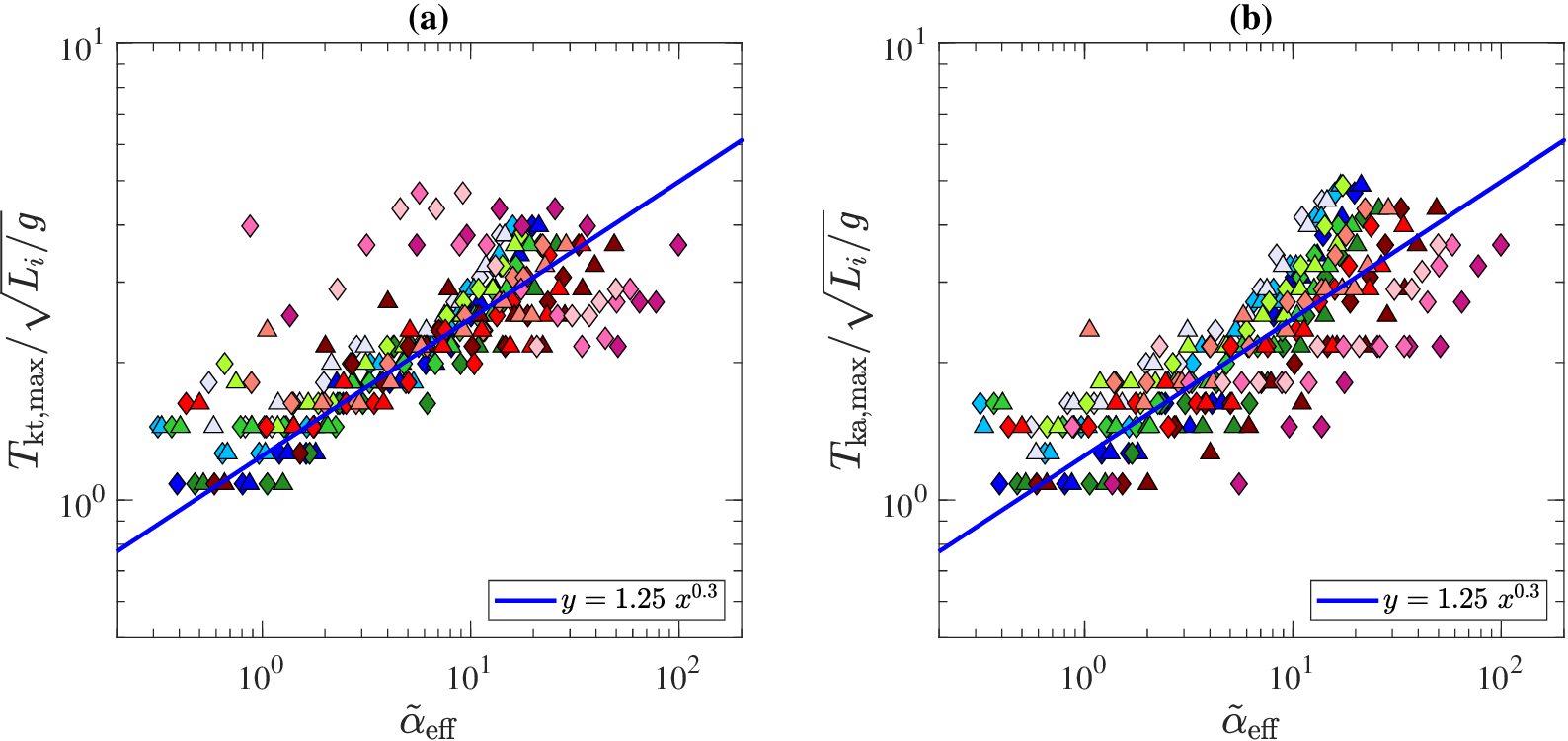}
  \caption{(a) The relationship between the dimensionless time when the system reaches its maximum translational kinetic energy, $\mathcal{T}_{\rm{kt,max}} = T_{\rm{kt,max}}/\sqrt{L_i/g}$, and $\Tilde{\alpha}_{\rm eff}$. (b) The relationship between the dimensionless time when the system reaches its maximum rotational kinetic energy, $\mathcal{T}_{\rm{ka,max}} = T_{\rm{ka,max}}/\sqrt{L_i/g}$, and $\Tilde{\alpha}_{\rm eff}$. Markers in this figure are the same with those in Figure \ref{fig:runout0.6}.}
\label{fig:TmaxEk}
\end{figure}

We have shown that changing the inclination angle has almost no influence on the maximum kinetic energy during column collapses. However, changing the inclination angle inevitably increases the available potential energy to be transformed into kinetic energy, which implies that the inclination angle is important to the time for a granular system to reach its peak kinetic energy. We focus on the time for the granular column collapse to reach its translational kinetic energy, $T_{\rm kt,max}$, and its rotational kinetic energy, $T_{\rm ka,max}$, and then normalize both $T_{\rm kt,max}$ and $T_{\rm ka,max}$ by $\sqrt{L_i/g}$. In Figure \ref{fig:TmaxEk}(a), we plot the dimensionless time for a system to reach its maximum translational kinetic energy, $\mathcal{T}_{\rm kt,max} = T_{\rm kt,max}/\sqrt{L_i/g}$, against $\Tilde{\alpha}_{\rm eff}$. For most simulation results ($\theta \leq 17.5^{\circ}$), $\mathcal{T}_{\rm kt,max}$ increases with the increase of $\Tilde{\alpha}_{\rm eff}$. $\mathcal{T}_{\rm kt,max}$ and $\Tilde{\alpha}_{\rm eff}$ have a strong power-law correlation. The scatteredness mainly comes from granular systems with $\theta = 20^{\circ}$, where increasing $\Tilde{\alpha}_{\rm eff}$ leads to a slight decrease in $\mathcal{T}_{\rm kt,max}$. From Figure \ref{fig:TmaxEk}(b), we see that $\mathcal{T}_{\rm ka,max}$ has similar scaling to $\mathcal{T}_{\rm kt,max}$, in that they both scale with $\Tilde{\alpha}_{\rm eff}^{0.3}$.

\subsection{Front velocity}

\begin{figure}
  \centering
  \includegraphics[scale=0.45]{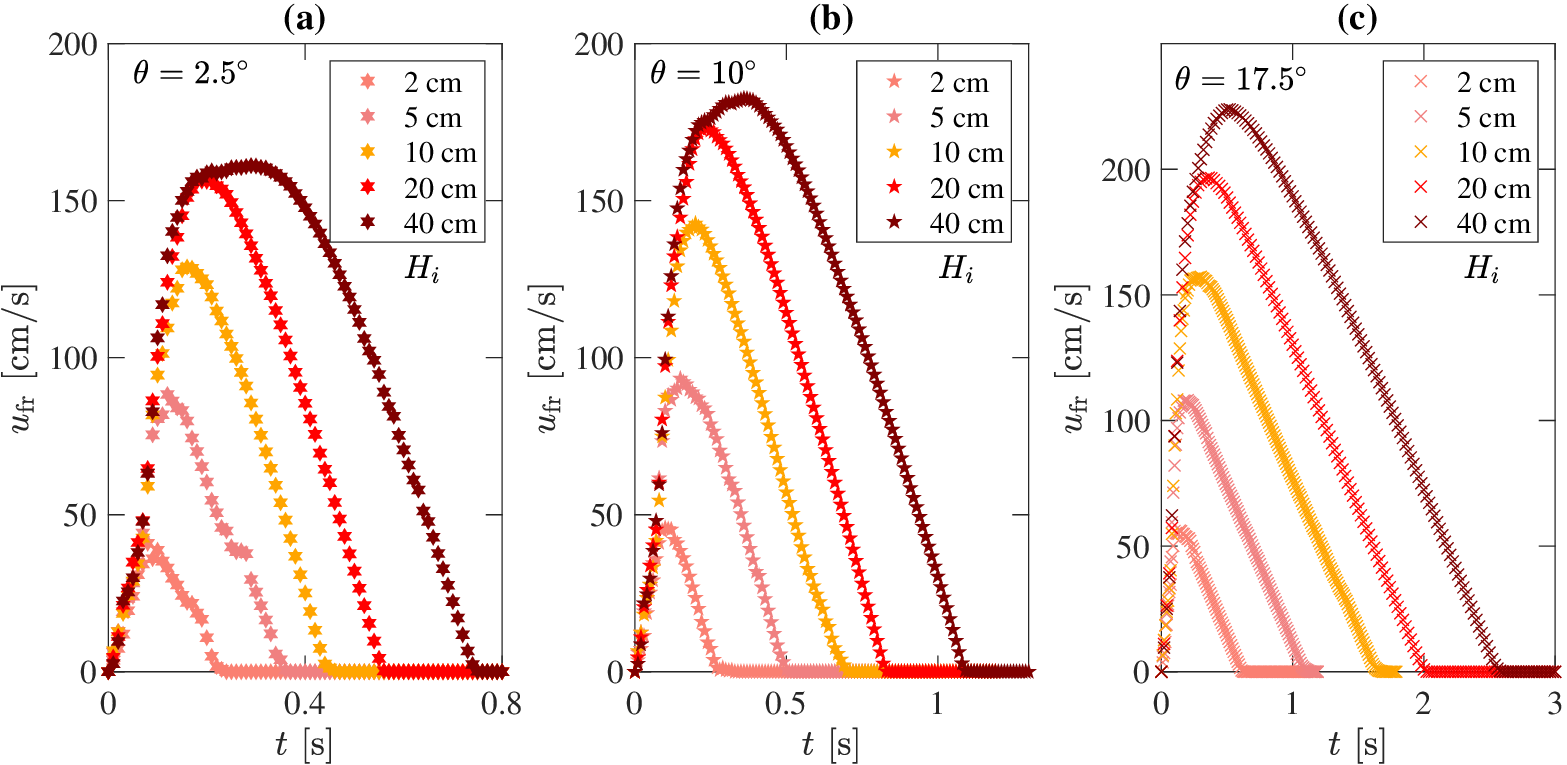}
  \caption{Time evolution of the front velocity for granular columns with (a) $\theta = 2.5^{\circ}$, (b) $\theta = 10^{\circ}$ and (c) $\theta = 17.5^{\circ}$. We set $\mu_p = 0.4$ in all three sets of simulation results.}
\label{fig:ufrTE}
\end{figure}

The maximum kinetic energy and the time for a system to reach its maximum kinetic energy measure the average capacity of the whole system to transform the potential energy into kinetic energy. If we regard a granular column collapse as a potential small scale landslide, we should also investigate its front velocity, $u_{\rm uf}$, since it directly links to the damage that a granular column collapse can cause to structures. For a granular column and at each time, we select a few particles located in the front and calculate their average velocity as the front velocity of this granular system. We plot the time evolution of $u_{\rm uf}$ for systems with $\theta = 2.5^{\circ}, 10^{\circ}$ and $17.5^{\circ}$ in Figure \ref{fig:ufrTE}. Similar to Figure \ref{fig:EkTime}, we only choose cases with $H_i =$ 2 cm, 5 cm, 10 cm, 20 cm and 40 cm to plot. Compared with Figure \ref{fig:EkTime}, we notice that the duration of $u_{\rm uf}$ is usually longer than that of the kinetic energy, especially when the initial aspect ratio is large. For instance, for granular systems with $\theta = 2.5^{\circ}$ and $H_i = 40$ cm, $u_{\rm fr}$ decays to 0 at $t \approx 0.75$ s, while $E_{\rm kt}$ declines to 0 before $t = 0.6$ s and $E_{\rm ka}$ decreases to 0 at $t \approx 0.65$. This is due to both $E_{\rm kt}$ and $E_{\rm ka}$ being averages of the whole granular system. After a granular system reaches its peak kinetic energy, most particles that are lagging behind stop moving, while only front particles continue to propagate, which results in a longer duration for $u_{\rm fr}$ than for the kinetic energy.

In Figure \ref{fig:EkTime}, for a granular column collapses, after reacking their peak kinetic energy, both $E_{\rm kt}$ and $E_{\rm ka}$ experience a exponential decay with respect to time. However, Figure \ref{fig:ufrTE} shows that the decay of the front velocity is approximately linear, instead of being almost exponential. Additionally, the linearity becomes more obvious when we tune the granular column to be taller and the inclined plane to be steeper. Both the initial aspect ratio and the inclination angle play important roles in determining the maximum front velocity and the collapse duration. In Figure \ref{fig:EkTime}, when $\theta = 2.5$ and $H_i = 10$ cm, the maximum front velocity is $u_{\rm fr,max}\approx 128.8$ at $T_{\rm fr,max} = 0.16$, and the front velocity lasts for 0.45 s. As we increase the inclination to $10^{\circ}$, $u_{\rm fr,max}$ only grows by 11\% and $T_{\rm fr,max}$ by 25\%, but the front velocity duration is increased by 57.8\%. Similar measurements occur when we increase the inclination angle from $10^{\circ}$ to $17.5^{\circ}$, while $u_{\rm fr,max}$, $T_{\rm fr,max}$ and the front velocity duration increase by 9.9\%, 50\% and 136.6\%, respectively. If we examine the results for systems with $H_i = 40$ cm, we can find similar behaviours. Then, we conclude that changing the inclination angle plays a more important role in determining the time-related information than that in quantifying the maximum front velocity. In other words, if a granular column collapse is considered as a landslide or a volcano-induced pyroclastic flow, a larger slope angle may not result in a heavier damage since the front velocity does not change much, it can certainly influence larger areas since the collapse duration is increased considerably.

\begin{figure}
  \centering
  \includegraphics[scale=0.45]{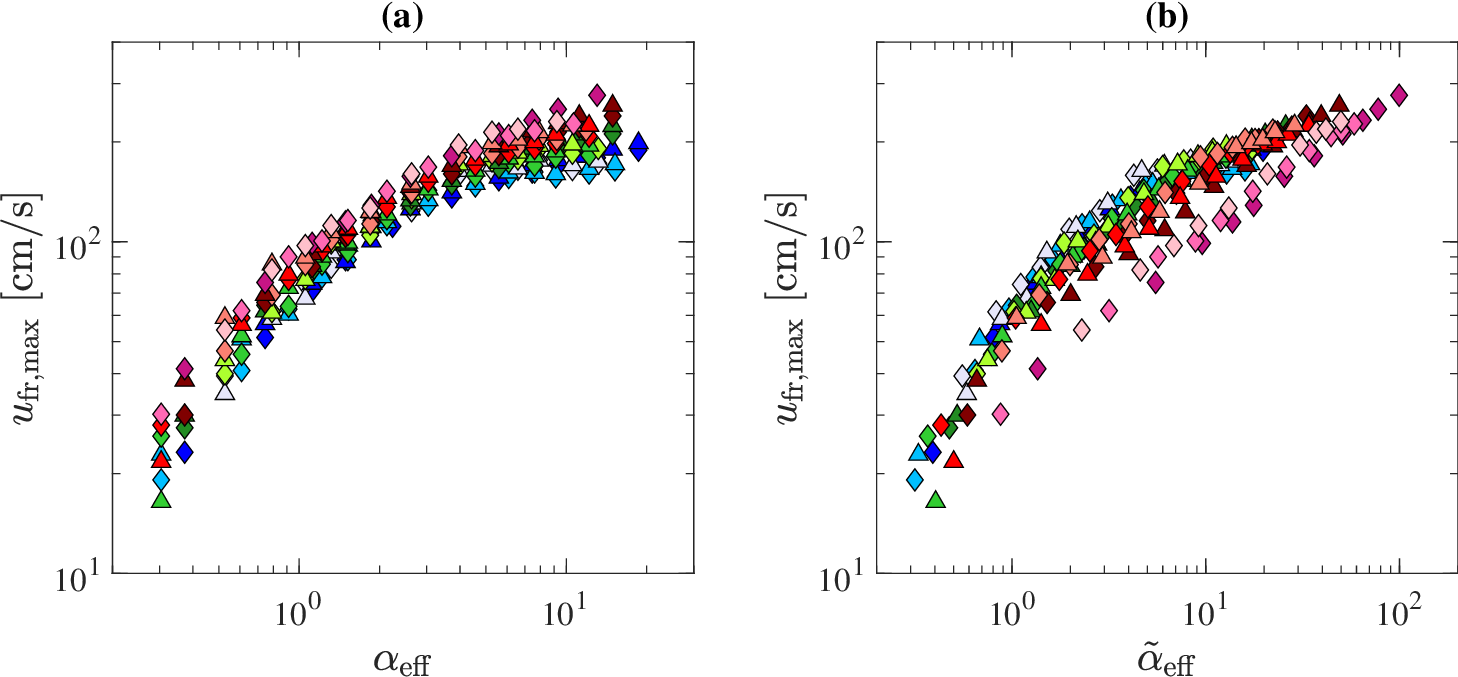}
  \caption{(a) The relationship between the maximum front velocity, $u_{\rm fr,max}$, and $\alpha_{\rm eff}$. (b) The relationship between the maximum front velocity, $u_{\rm fr,max}$, and $\Tilde{\alpha}_{\rm eff}$. Markers in this figure are the same as those in Figure \ref{fig:runout0.6}.}
\label{fig:ufr_n_alpha}
\end{figure}

\begin{figure}
  \centering
  \includegraphics[scale=0.45]{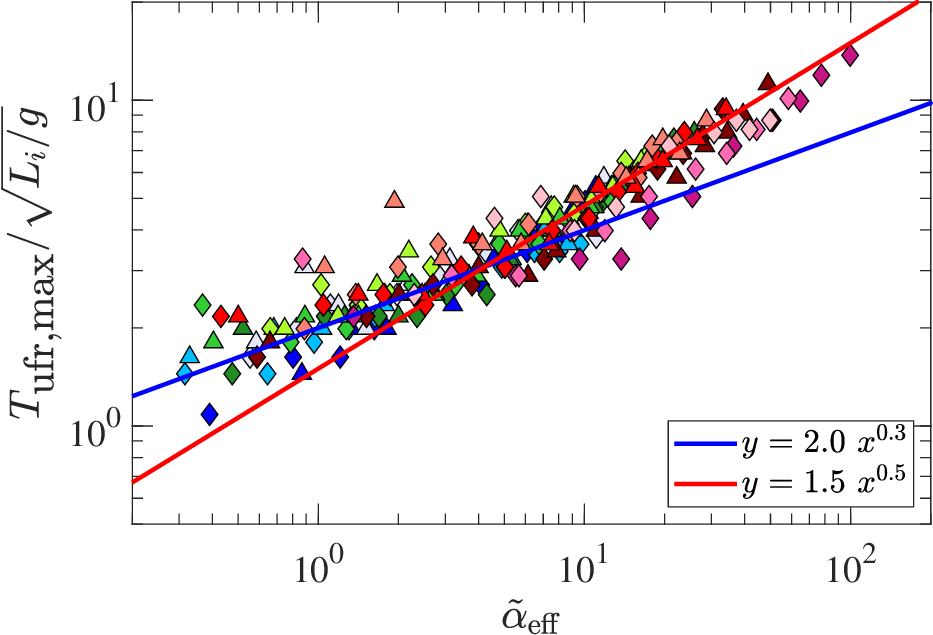}
  \caption{The relationship between the time when a system reaches its maximum front velocity, $T_{\rm ufr,max}$, and $\Tilde{\alpha}_{\rm ufr,max}$. The $y$ axis is normalized by $\sqrt{L_i/g}$. Markers in this figure are the same as those in Figure \ref{fig:runout0.6}.}
\label{fig:Tufrmax}
\end{figure}

Similarly, we expect that ${\alpha}_{\rm eff}$ would have better performance than $\Tilde{\alpha}_{\rm eff}$ in its relationship with $u_{\rm fr,max}$. Figure \ref{fig:ufr_n_alpha} confirms our expectation to show that ${\alpha}_{\rm eff}$ and $u_{\rm fr,max}$ have a clear correlation, and as we increase the inclination angle, $u_{\rm fr,max}$ is only increased slightly. However, different from the relationship between the kinetic energy and $\Tilde{\alpha}_{\rm eff}$ where the scatteredness is pervasive, the $u_{\rm fr,,ax}\sim\Tilde{\alpha}_{\rm eff}$ relationship almost concentrates on a master curve, if we exclude data from granular column collapses with $\theta = 17.5^{\circ}$ and $20^{\circ}$. Based on the maximum front velocity, it seems that granular column collapses can be classified into two types: (1) when $\Tilde{\alpha}_{\rm eff}\lessapprox 4$, $u_{\rm fr,max}$ increases rapidly as we increase $\Tilde{\alpha}_{\rm eff}$; (2) when $\Tilde{\alpha}_{\rm eff}\gtrapprox 4$, the increase of $u_{\rm fr,max}$ becomes much slower with respect to the increase of $\Tilde{\alpha}_{\rm eff}$. We then plot the relationship between $\mathcal{T}_{\rm ufr,max} \equiv T_{\rm ufr,max}/\sqrt{L_i/g}$ and $\Tilde{\alpha_{\rm eff}}$ in Figure \ref{fig:Tufrmax}. The relationship between $\mathcal{T}_{\rm ufr,max}$ and $\Tilde{\alpha}_{\rm eff}$ is two-stage exponential with a dividing point at $\Tilde{\alpha}_{\rm eff} \approx 4$ with
\begin{equation} \label{eq:Tufrmax}
  \mathcal{T}_{\rm ufr,max} \equiv \frac{T_{\rm ufr,max}}{\sqrt{L_i/g}} \approx \left\{
    \begin{array}{ll}
      2.0 \Tilde{\alpha}_{\rm eff}^{0.3}, & \Tilde{\alpha}_{\rm eff}\lessapprox 4 \\[6pt]
      1.5 \Tilde{\alpha}_{\rm eff}^{0.5},         & \Tilde{\alpha}_{\rm eff}\gtrapprox 4.
    \end{array} \right.
\end{equation}

We note that the dividing point in Equation \ref{eq:Tufrmax} is the same as that in Figure \ref{fig:ufr_n_alpha}(b). In the $u_{\rm fr,max}\sim\Tilde{\alpha}_{\rm eff}$ relationship, as we increase $\Tilde{\alpha}_{\rm eff}$, the slope shown on a log-log coordinate system begins with a large value and then decreases. However, in the $\mathcal{T}_{\rm ufr,max}\sim\Tilde{\alpha}_{\rm eff}$ relationship, the slope of the power-law relation is smaller for systems with $\Tilde{\alpha}_{\rm eff}\lessapprox 4$ than that for systems with larger $\Tilde{\alpha}_{\rm eff}$. In previous research \citep{man2023bifriction}, we discovered that, for horizontal granular column collapses, the time at which a granular column collapse reaches its peak kinetic energy scales with $\alpha_{\rm eff}^{0.5}$, which is the same as results from \citet{lube2004axisymmetric}. We then expect that, for granular column collapses on an inclined plane, the scaling of $\mathcal{T}_{\rm kt,max}$, $\mathcal{T}_{\rm ka,max}$ and $\mathcal{T}_{\rm ufr,max}$ should behave similarly. However, for both kinetic energies, their power-law exponent is equal to 0.3, which is smaller than that for systems on horizontal planes. This indicates that both $\mathcal{T}_{\rm kt,max}$ and $\mathcal{T}_{\rm ka,max}$ increase slowly with the increase of $\Tilde{\alpha}_{\rm eff}$. It is often faster for a granular system to reach its peak kinetic energy when it is on a slope than that on a horizontal plane. Figure \ref{fig:Tufrmax} shows that $\mathcal{T}_{\rm ufr,max}$ also scales with $\Tilde{\alpha}_{\rm eff}^{0.3}$ when $\Tilde{\alpha}_{\rm eff} \lessapprox 4$, but the maximum front velocity for systems with inclinations usually comes later than a system on a horizontal plane, which contradicts the fact that column collapses on inclined planes reach peak kinetic energy earlier than those on horizontal planes. For a horizontal granular column collapse, its peak front velocity usually comes with the maximum kinetic energy, whereas for inclined collapses, there exists a clear gap between  $T_{\rm kt,max}$ and $T_{\rm ufr,max}$, which may result from the flow front being subjected to smaller frictional effects due to the inclination, but detailed study is still needed in future works.

\subsection{Terminal time}
\label{sec:Tf}
\begin{figure}
  \centerline{\includegraphics[scale=0.45]{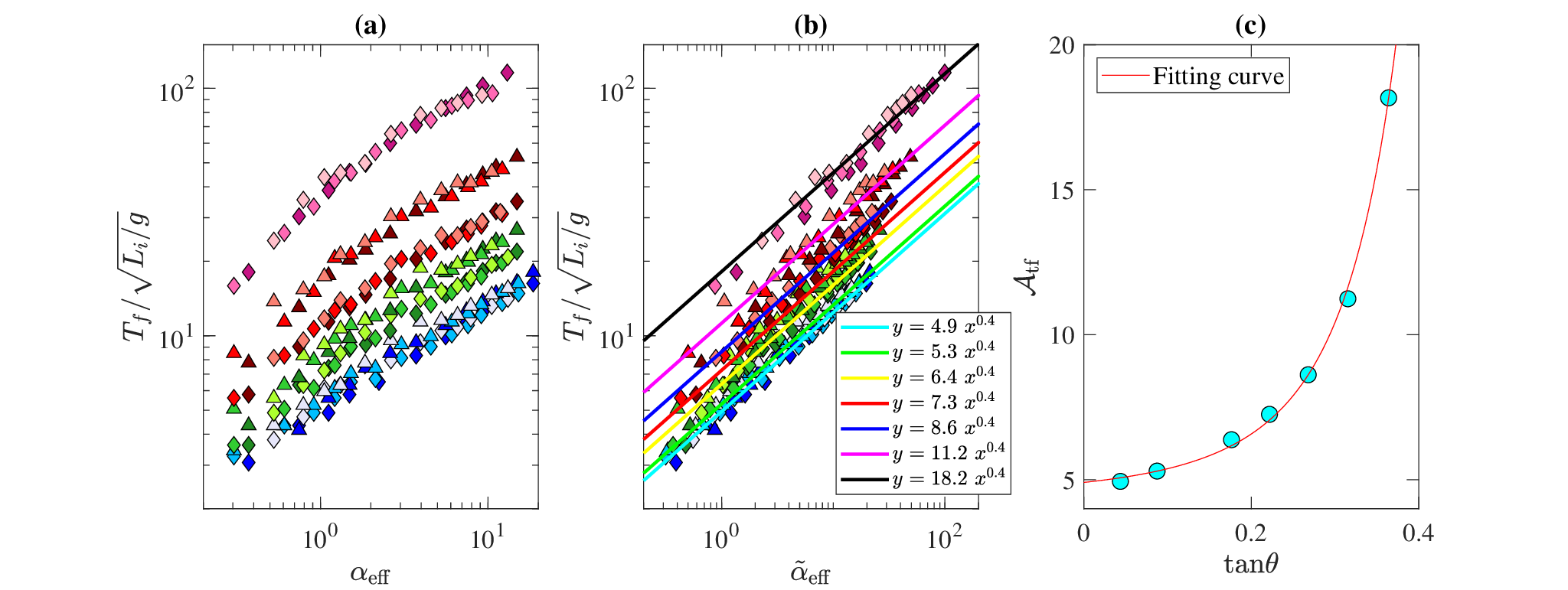}}
  \caption{(a) The relationship between the dimensionless collapse duration, $\mathcal{T}_f \equiv T_f/\sqrt{L_i/g}$, and $\alpha_{\rm eff}$. (b) The relationship between $\mathcal{T}_f$ and the inclined effective aspect ratio, $\Tilde{\alpha}_{\rm eff}$. Fitted curves follow power-law scalings with $\mathcal{T}_{f} = \mathcal{A}_{\rm tf}\cdot\Tilde{\alpha}_{\rm eff}^{\zeta}$, where $\zeta$ and $\mathcal{A}_{\rm tf}$ are fitted parameters. (c) The relationship between $\mathcal{A}_{\rm tf}$ and $\tan\theta$. Markers in Figures (a) and (b) are the same as those in Figure \ref{fig:runout0.6}}
\label{fig:Tfplot}
\end{figure}

We further investigate how much time it takes for a granular column collapse to come to rest, which can also be regarded as the collapse duration. For a granular column with the same initial aspect ratio, as we increase the inclination angle, the run-out distance will increase accordingly, which may result in a longer propagation period and a larger value of the terminal time, $T_f$. In this work, we define $T_f$ based on the time evolution of the front velocity and regard the time when the front velocity diminishes as the terminal time for the collapse. Then, a dimensionless terminal time can be defined as $\mathcal{T}_{f}\equiv T_f/\sqrt{L_i/g}$. Both \citet{lube2004axisymmetric} and \citet{lube2005collapses} stated that, based on dimensional analysis, $T_f$ must scale with $(L_i/g)^{0.5}\psi(\alpha)$, where $\psi(\alpha)$ is a function of the initial aspect ratio. \citet{man2023bifriction} then concluded that, for systems with different frictional coefficients, $\psi(\alpha)$ should be modified to $\psi(\alpha_{\rm eff})$, and argued that $\psi(\alpha_{\rm eff}) = \kappa_t\alpha_{\rm eff}^{0.5}$, where $\kappa_t$ is a constant, is the best fit to the simulation results. In this work, we further hypothesize that $\phi(\cdot)$ is a function of $\Tilde{\alpha}_{\rm eff}$ since the new dimensionless number works well for $\mathcal{T}_{\rm kt,max}$, $\mathcal{T}_{\rm ka,max}$ and $\mathcal{T}_{\rm ufr,max}$. 

We start with the relationship between $\mathcal{T}_{f}\equiv T_f/\sqrt{L_i/g}$ and $\alpha_{\rm eff}$ shown in Figure \ref{fig:Tfplot}(a). $\mathcal{T}_f$ is highly $\theta$-dependent, and $\mathcal{T}_f$ increases with an increase in the inclination angle. For each set of simulations with the same $\theta$, $\alpha_{\rm eff}$ successfully combines simulation results with different frictional properties. However, contrary to our expectation that increasing $\theta$ only shifts the $\mathcal{T}_f\sim\alpha_{\rm eff}$ curve upward, changing $\theta$ also modifies the shape of the $\mathcal{T}_f\sim\alpha_{\rm eff}$ curve. For systems with $\theta = 2.5^{\circ}$ and $5^{\circ}$, $\mathcal{T}_f$ and $\alpha_{\rm eff}$ have a power-law relationship. For systems with larger inclination angles, the $\mathcal{T}_f\sim\alpha_{\rm eff}$ relationship seems to have two power-law relations divided by critical $\alpha_{\rm eff}$'s, which are also dependent on the inclination angle. We then plot $\mathcal{T}_f$ against the inclined effective aspect ratio, $\Tilde{\alpha}_{\rm eff}$, in Figure \ref{fig:Tfplot}(b), which shows clearly that $\mathcal{T}_f$ and $\Tilde{\alpha}_{\rm eff}$ follow power-law relationships with the same power-law exponent. Their relationship can be expressed using the following equation that
\begin{equation} \label{eq:Tf}
    \begin{split}
        \mathcal{T}_{f} \equiv \frac{T_f}{\sqrt{L_i/g}} = \mathcal{A}_{\rm tf}\cdot\Tilde{\alpha}_{\rm eff}^{\zeta}\ ,
    \end{split}
\end{equation}
where $\zeta = 0.4$ is a constant, and $\mathcal{A}_{\rm tf}$ is a fitted scalar that is only dependent on the inclination angle. The relationship between $\mathcal{A}_{\rm tf}$ and $\theta$ is shown in Figure \ref{fig:Tfplot}(c), where $\mathcal{A}_{\rm tf}$ increases almost exponentially with respect to the increase of $\tan{\theta}$. We fit the $\mathcal{A}_{\rm tf}\sim\tan{\theta}$ relationship with the following equation,
\begin{equation} \label{eq:Atf}
    \begin{split}
        \mathcal{A}_{\rm tf} = A_o + \epsilon_1\cdot\exp\left( \frac{\Theta_f}{s_t-\tan\theta} \right)\ , \tan\theta\leq s_t\ ,
    \end{split}
\end{equation}
where $A_o \approx 4.328$, $\epsilon_1\approx 3.76\times 10^{-3}$, $\Theta_f \approx 4.74$ and $s_t \approx 0.94$ are fitted parameters. The R-squared of the fitting curve is approximately 0.997. This indicates that, when $\tan\theta$ is approaching $s_t$, $\mathcal{A}_{\rm tf}$ is inevitably reaching infinity. We note that, in this work, the frictional coefficient  between the inclined plane and particles is $\mu_w = 0.4$, yet the fitted $s_t$ is much larger than $\mu_w$, which implies that an inclined plane with slope larger than $\arctan(\mu_w)$ can still hold Voronoi-based grains. This may be because that Voronoi-based particles innately have rolling resistances due to their random, non-spherical, angular shapes.

\begin{figure}
  \centerline{\includegraphics[scale=0.45]{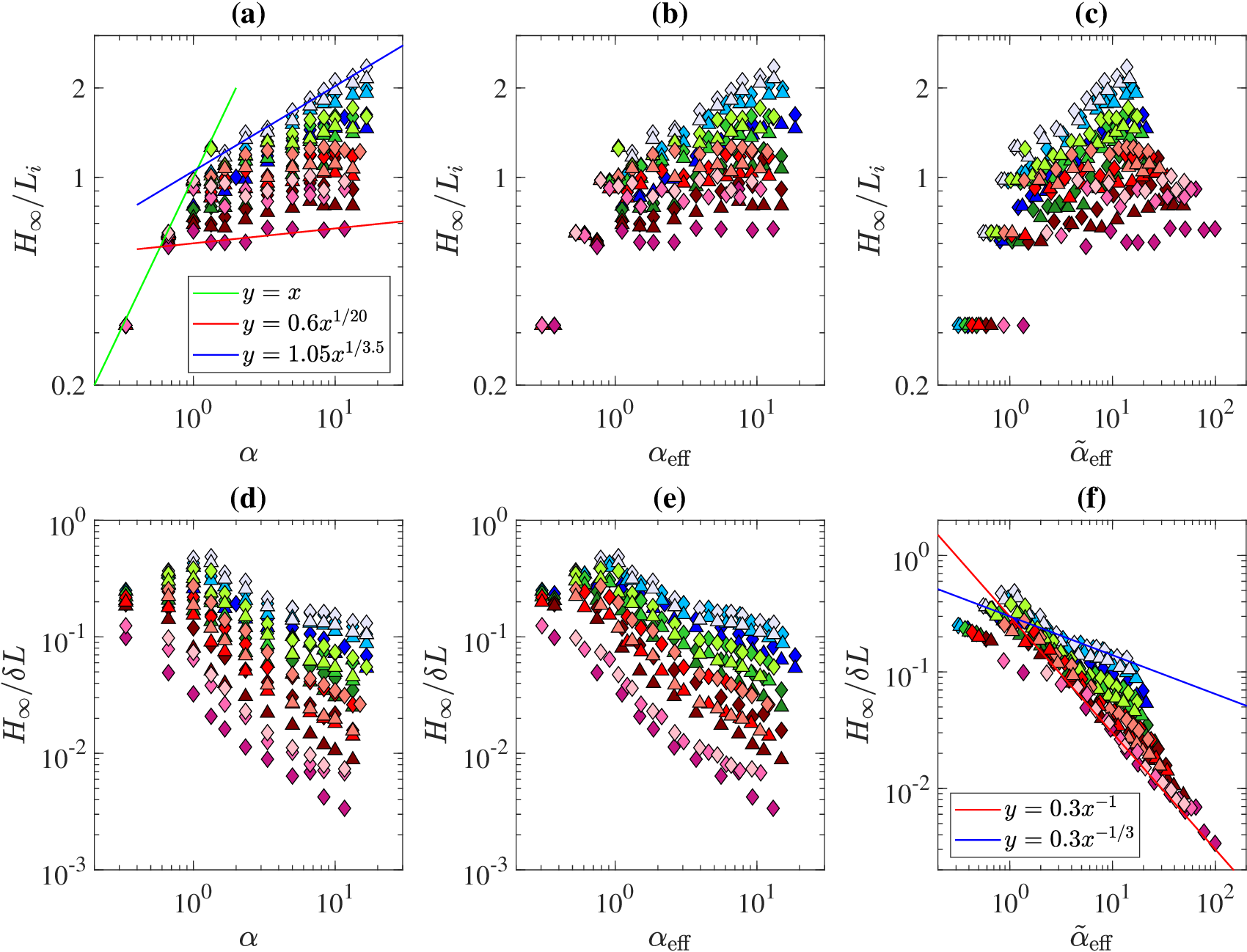}}
  \caption{(a-c) show relationships between $H_{\infty}/L_i$ and $\alpha$, $\alpha_{\rm eff}$ and $\Tilde{\alpha}_{\rm eff}$, respectively. (d-f) plot relationships between $H_{\infty}/\delta L$ and $\alpha$, $\alpha_{\rm eff}$ and $\Tilde{\alpha}_{\rm eff}$, respectively. Markers are the same as those in Figure \ref{fig:runout0.6}}
\label{fig:Hfplot1}
\end{figure}

\section{Final deposition height}
\label{sec:height}

For granular column collapses, longer run-out distances often result in shorter final deposition heights, $H_{\infty}$. In the previous study \citep{man2023bifriction}, we showed the complexity of the relationship between $H_{\infty}$ and $\alpha_{\rm eff}$ of systems with different inter-particle frictional properties. However, the complexity was somewhat overcome as we, instead of focusing only on $H_{\infty}$, analyzed the volume of the deposition cone, $\mathcal{V}_{\rm cone}$, and the ratio between $\mathcal{V}_{\rm cone}$ and the initial column volume $\mathcal{V}_{\rm init}$. In this work, we first investigate the relationship between $H_{\infty}/L_i$ and dimensionless numbers as shown in Figure \ref{fig:Hfplot1}(a-c).

The relationship between $H_{\infty}/L_i$ and $\alpha$ exhibits power-law characteristics of $H_{\infty}/L_i\sim\alpha^{\xi}$ as shown in Figure \ref{fig:Hfplot1}(a). For granular columns with the same frictional property and the same inclination angle but different initial column heights, when $\alpha$ is below a threshold $\alpha_{\rm hc}$, $H_{\infty}/L_i$ always scales with $\alpha$ and $\xi = 1$ (the green line in Figure \ref{fig:Hfplot1}(a)), but when $\alpha$ is larger than a threshold $\alpha_{\rm hc}$, $H_{\infty}/L_i$ scales with $\alpha^{\xi}$ and $\xi$ is much smaller than 1. When $\alpha>\alpha_{\rm hc}$, as we increase the inclination angle from 2.5$^{\circ}$ to 20$^{\circ}$ and the inter-particle frictional coefficient from 0.6 to 0.2, the power-law exponent, $\xi$, decreases from $\approx 1/3.5$ to $\approx 1/20$, which indicates that it is difficult for granular columns with larger inclination angles to sustain a larger deposition height as particles tend to flow further on planes with larger $\theta$. As expected from, and similar to our previous study \citep{man2023bifriction}, changing the $x$ axis from $\alpha$ to $\alpha_{\rm eff}$  or $\Tilde{\alpha}_{\rm eff}$ (Figure \ref{fig:Hfplot1}(b) and (c)), instead of solving the discreteness of simulation data, further increases the pronounced scattering, which leads us to combine deposition height with the run-out behaviour.

In Figure \ref{fig:Hfplot1}(d), the ratio between $H_{\infty}$ and $\delta L$ is plotted against the initial aspect ratio $\alpha$, which presents some interesting phenomena. On one hand, when keeping $\alpha$ constant, both increasing the inclination angle and decreasing the inter-particle frictional coefficient result in a decrease in $H_{\infty}/\delta L$. This is due to that both increasing $\theta$ and decreasing $\mu_p$ make the granular system easier to flow and lead to larger mobility, larger run-out distance and shorter final deposition height. On the other hand, while keeping $\theta$ and $\mu_p$ constant, the relationship between $H_{\infty}/\delta L$ and $\alpha$ varies from case to case. When $\theta\leq 15^{\circ}$, as we increase $\alpha$ from 0.3 to 15, $H_{\infty}/\delta L$ first increase and then decrease. The transitional point is approximately at $\alpha = 1.4$. However, when $\theta \geq 17.5^{\circ}$, the $H_{\infty}/\delta L \sim \alpha$ relationship shows pronounced monotonic decreasing pattern. Changing the $x$ axis to $\alpha_{\rm eff}$ in Figure \ref{fig:Hfplot1}(e), although it cannot reflect the influence of the inclination angle, does help unify the influence of frictional properties that the $H_{\infty}/\delta L\sim \alpha_{\rm eff}$ relationship of systems with different frictional coefficients but the same inclination angle seems to collapse onto one master curve.

\begin{figure}
  \centerline{\includegraphics[scale=0.45]{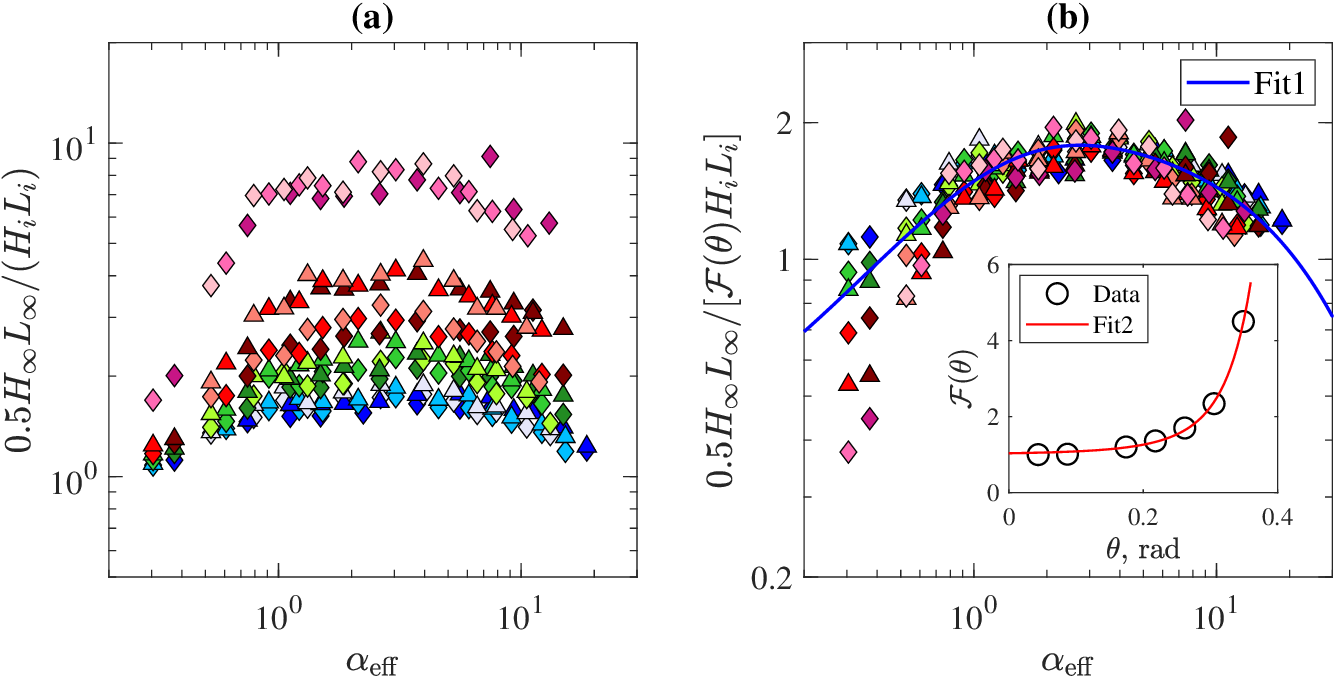}}
  \caption{Markers are the same as those in Figure \ref{fig:runout0.6}}
\label{fig:Hfplot2}
\end{figure}

Plotting $H_{\infty}/\delta L$ against $\Tilde{\alpha}_{\rm eff}$, in some sense, converges simulation results with different inclination angles, yet systems with different $\theta$ still has different power-law scaling exponents. When $\theta = 2.5^{\circ}$, $H_{\infty}/\delta L$ scales with $(\Tilde{\alpha}_{\rm eff})^{-1}$, whereas $H_{\infty}/\delta L$ scales with $(\Tilde{\alpha}_{\rm eff})^{-1/3}$ when $\theta = 20^{\circ}$. Learning from the previous research \citep{man2023bifriction}, we switch to analyze the ratio between final and initial column area in the $x-z$ plane (the width of the granular column before or after the collapse remains the same). The initial column area and the approximation of the final area (ignoring the complexity of the final deposition pattern) are listed as follows that
\begin{subequations}
    \begin{align}
        \mathcal{A}_{\rm init} = H_{i}L_{i}\ , \\
        \mathcal{A}_{\rm fin} = 0.5H_{\infty}L_{\infty}\ ,
    \end{align}
\end{subequations}
where $\mathcal{A}_{\rm init}$ is the initial column area in the $x-z$ plane and $\mathcal{A}_{\rm fin}$ is an approximation of the final deposition area. On one hand, based on our understanding of deposition patterns of granular column collapses reported in \citet{man2021deposition}, with the increase of $\theta$, the granular system becomes more and more fluid-like, which results in a longer run-out distance and a more complex deposition pattern. The change in deposition morphology leads to an exaggeration of the deposition area with the approximation of $\mathcal{A}_{\rm fin} = 0.5H_{\infty}L_{\infty}$. Thus, we expect that, as we increase $\theta$, the ratio between  $\mathcal{A}_{\rm fin}$ and $\mathcal{A}_{\rm init}$ will also increase. On the other hand, similar to \citet{man2023bifriction}, while keeping $\theta$ constant, increasing $\alpha_{\rm eff}$ may result in an increase-decrease relationship between $\mathcal{A}_{\rm fin}/\mathcal{A}_{\rm init}$ and $\alpha_{\rm eff}$.

Plotting the relationship between $\mathcal{A}_{\rm fin}/\mathcal{A}_{\rm init} = 0.5H_{\infty}L_{\infty}/H_{i}L_{i}$ and $\alpha_{\rm eff}$ in Figure \ref{fig:Hfplot2}(a) reveals that our expectation is only partially realized, in that increasing $\theta$ surely leads to an increase in $0.5H_{\infty}L_{\infty}/H_{i}L_{i}$. However, when we keep $\theta$ constant while varying $\alpha_{\rm eff}$, $0.5H_{\infty}L_{\infty}/H_{i}L_{i}$ increases with $\alpha_{\rm eff}$ when $\alpha_{\rm eff} \lessapprox 1$, $0.5H_{\infty}L_{\infty}/H_{i}L_{i}$ decreases with $\alpha_{\rm eff}$ when $\alpha_{\rm eff} \gtrapprox 5$, but $0.5H_{\infty}L_{\infty}/H_{i}L_{i}$ remain almost at a plateau within the interval of $\alpha_{\rm eff}\in (1, 5)$. We hypothesize that the plateau is caused by the change of $\theta$, and we perform a separation of variables so that $\mathcal{A}_{\rm fin}/\mathcal{A}_{\rm init}$ can be written as a multiplication of a function of $\theta$ and a function of $\alpha_{\rm eff}$ and fit the influence of $\alpha_{\rm eff}$ with a double-exponential equation, so that
\begin{equation} \label{eq:Hf}
    \begin{split}
        \frac{\mathcal{A}_{\rm fin}}{\mathcal{A}_{\rm init}} = \frac{0.5H_{\infty}L_{\infty}}{H_{i}L_{i}} = \mathcal{F}(\theta) \left[ a_1 \exp(-b_1\alpha_{\rm eff}) - a_2\exp(-b_2\alpha_{\rm eff})  \right]\ ,
    \end{split}
\end{equation}
where $a_1 \approx 2$, $b_1 \approx 0.0327$, $a_2 \approx 1.68$ and $b_2 \approx 1.32$ are fitting parameters, and $\mathcal{F}(\theta)$ is a function of $\theta$, which determines the level of plateau in the relationship between $\mathcal{A}_{\rm fin}/\mathcal{A}_{\rm init}$ and $\alpha_{\rm eff}$. Figure \ref{fig:Hfplot2}(b) confirms our hypothesis that $0.5H_{\infty}L_{\infty}/[\mathcal{F}(\theta)H_{i}L_{i}]$ can be expressed by the function of $\theta$ shown in Eq. \ref{eq:Hf}. When $\alpha_{\rm eff}\lessapprox 1$, the data points are scattered, but the fitting curve of Eq. \ref{eq:Hf} performs well for systems with $\alpha_{\rm eff}\gtrapprox 1$. We can also obtain $\mathcal{F}(\theta)$ as plotted in the inset of Figure \ref{fig:Hfplot2}(b), which is analogous to the relationship between $\mathcal{A}_{\rm tf}$ and $\theta$ shown in Section \ref{sec:Tf}, that
\begin{equation} \label{eq:Ftheta}
    \begin{split}
        \mathcal{F}(\theta) = 1 + \epsilon_2\exp\left( \frac{\Theta_h}{s_t - \tan\theta}\right)\ ,
    \end{split}
\end{equation}
where $\epsilon_2\approx 2.435\times 10^{-5}$ and $\Theta_h\approx 6.85$. We note that, when $\tan\theta = s_t$, $\mathcal{F}(\theta)$ is approaching $\infty$, which is the same as the relationship between $\mathcal{A}_{\rm tf}$ and $\theta$.

\section{Influence of initial solid fractions}
\label{sec:solidfrac}

\begin{figure}
  \centerline{\includegraphics[scale=0.45]{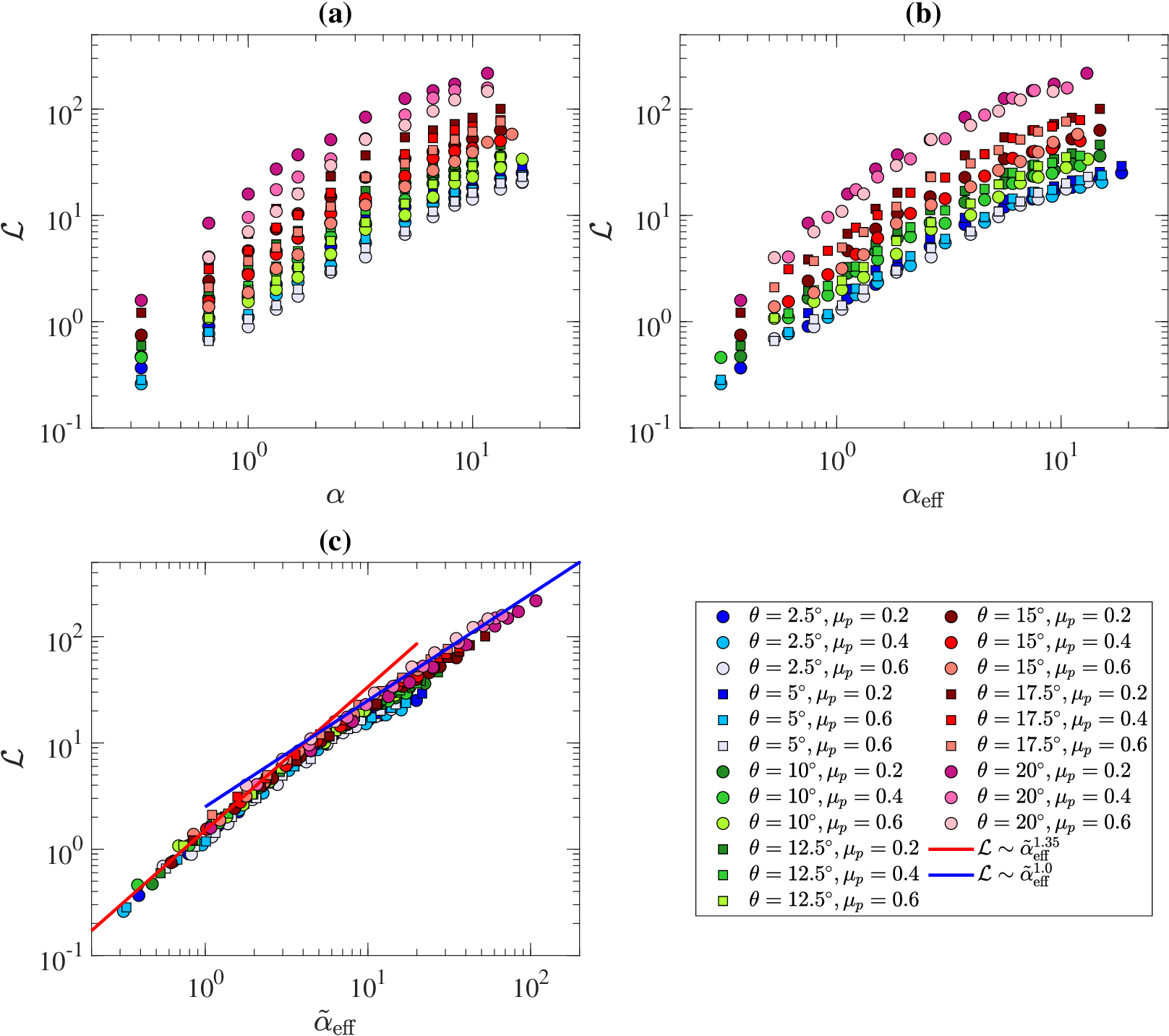}}
  \caption{Relative horizontal run-out distance of systems with $\phi_{\rm init} = 0.8$ plotted against (a) initial aspect ratios, $\alpha$, (b) effective aspect ratios, $\alpha_{\rm{eff}}$, and (c) inclined effective aspect ratio, $\Tilde{\alpha}_{\rm{eff}}$, for 21 different sets of simulations. The red curve represents the fitted relationship of $\mathcal{L}\sim\Tilde{\alpha}_{\rm eff}^{1.35}$ and the blue curve denotes the fitting of $\mathcal{L}\sim\Tilde{\alpha}_{\rm eff}$}
\label{fig:runout0.8}
\end{figure}

The phenomena listed in previous sections are acquired from our study of granular systems with $\phi_{\rm init} = 0.6$. However, the initial solid fraction often plays an important role in determining the macroscopic behaviour of granular flows, especially for systems in subaqueous environments \citep{Pailha2008}. Even for dry granular systems, changing initial solid fraction can evidently affect the dynamical behaviour \citep{man2023bifriction}. Thus, we perform another set of simulations for granular column collapses with $\phi_{\rm init} = 0.8$ and investigate how changing the initial solid fraction influences the run-out behaviour, the collapse duration and the deposition height. For granular columns with $\phi_{\rm init} = 0.8$, the column collapse may result in a dilation process due to the shearing effect, which can lead to a longer run-out distance.

\begin{figure}
  \centerline{\includegraphics[scale=0.45]{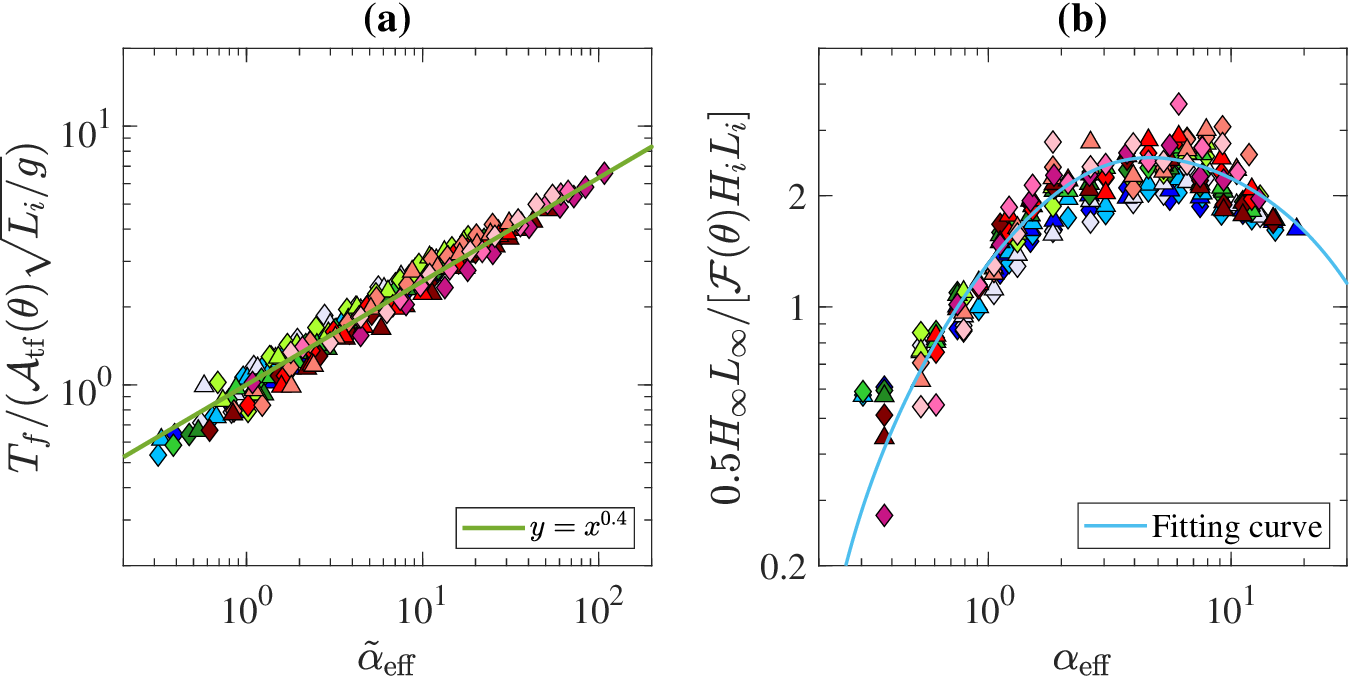}}
  \caption{(a) The relationship between $T_f/(\mathcal{A}_{\rm tf}\sqrt{L_i/g})$ and $\Tilde{\alpha}_{\rm eff}$ for granular columns with initial solid fraction $\phi_{\rm init} = 0.8$, where $\mathcal{A}_{\rm tf}$ is calculated with Eq. \ref{eq:Atf}. (b) The relationship between $0.5H_{\infty}L_{\infty}/[\mathcal{F}(\theta)H_i L_i]$ and $\alpha_{\rm eff}$ for granular column collapses with $\phi_{\rm init} = 0.8$. Markers are the same as those in Figure \ref{fig:runout0.8}}
\label{fig:TfHflargeSF}
\end{figure}

In Figure \ref{fig:runout0.8}, we plot relationships between $\mathcal{L}$ and $\alpha$, $\alpha_{\rm eff}$ and $\Tilde{\alpha}_{\rm eff}$, respectively. Contrary to our expectation that increasing the solid fraction leads to pronouncedly larger $\mathcal{L}$, simulation results for systems with $\phi_{\rm init}=0.8$ do not differ much from those for systems with $\phi_{\rm init} = 0.6$. We note that, when we plot $\mathcal{L}$ against $\Tilde{\alpha}_{\rm eff}$ in Figure \ref{fig:runout0.8}(c), simulation data collapse better than those of systems with $\phi_{\rm init} = 0.6$. The only difference between the two plots shown in Figures \ref{fig:runout0.6} and \ref{fig:runout0.8} brought about by raising $\phi_{\rm init}$ from 0.6 to 0.8 is that when $\Tilde{\alpha}_{\rm eff}>10$ and $\theta\leq 5^{\circ}$, the run-out distance of systems with $\phi_{\rm init} = 0.8$ is larger than that of systems with $\phi_{\rm init} = 0.6$. This leads us to the conclusion that, for dry granular columns with initially stable structures, the initial solid fraction is less significant compare to other parameters, if the granular column collapse occurs on an inclined plane. The new fitting curve shown as the light blue curve in Figure \ref{fig:TfHflargeSF}(b) can also be written as a double-exponential equation with the same functional form as Equation \ref{eq:Ftheta} but with different fitting parameters ($a_1\approx 3.09$, $b_1\approx 0.0327$, $a_2\approx 3.45$ and $b_2\approx 0.72$)

Based on the influence of the initial solid fraction on the run-out behaviour of granular column collapses on inclined planes, we can expect that the collapse duration, $T_f$, of systems with $\phi_{\rm init}=0.8$ behaves similar as that of granular columns with $\phi_{\rm init}=0.6$. As we plot in Figure \ref{fig:TfHflargeSF}(a) the relationship between $T_f/(\mathcal{A}_{\rm tf}\sqrt{L_i/g})$ and $\Tilde{\alpha}_{\rm eff}$, where $\mathcal{A}_{\rm tf}$ is calculated with Eq. \ref{eq:Atf}, the result confirms our expectation with a good collapse of all the data onto a master curve of $T_f/(\mathcal{A}_{\rm tf}\sqrt{L_i/g}) = \Tilde{\alpha}_{\rm eff}^{0.4}$, which is exactly the same as what we have obtained in equations \ref{eq:Tf} and \ref{eq:Atf}. This indicates that, as long as the initial granular packing is stable under self-weight, changing the initial solid fraction has no influence on the collapse duration. However, the initial solid fraction does play an important role in the final deposition height of a collapsed granular column since the initial packing structure influences the initial stability and the initial failure criterion of granular columns. Consequently, larger initial solid fraction often indicates that fewer particles participate in the granular avalanche and leads to higher final deposition height.

To examine the feasibility of equations \ref{eq:Hf} and \ref{eq:Ftheta} for granular columns with $\phi_{\rm init} = 0.8$, we plot the relationship between $0.5H_{\infty}L_{\infty}/[\mathcal{F}(\theta)H_i L_i]$ and $\alpha_{\rm eff}$ in Figure \ref{fig:TfHflargeSF}(b), where $\mathcal{F}(\theta)$ is calculated using equation \ref{eq:Ftheta} with the exact same parameters. With the renormalization of $\mathcal{F}(\theta)$, all the simulation data collapse onto one master curve, but this master curve is evidently different from that shown in Figure \ref{fig:Hfplot2}(b). In Figure \ref{fig:Hfplot2}(b), $0.5H_{\infty}L_{\infty}/[\mathcal{F}(\theta)H_i L_i]$ agrees with a double-exponential equation and approximately varies from 0.5 to 2 as $\alpha_{\rm eff}$ changes from 0.3 to 20. However, as shown in Figure \ref{fig:TfHflargeSF}(b), for systems with larger initial solid fraction, $0.5H_{\infty}L_{\infty}/[\mathcal{F}(\theta)H_i L_i]$ varies from $0.4$ to $\approx 3.5$, which confirms that granular columns with larger $\phi_{\rm init}$ tend to sustain taller final deposition heights than systems with small initial solid fractions. We hypothesize that the change of the $0.5H_{\infty}L_{\infty}/[\mathcal{F}(\theta)H_i L_i] \sim \alpha_{\rm eff}$ relationship results from the $\phi_{\rm init}-$induced change of yielding behaviour of the granular packing, which will be further investigated in future studies.

\section{Further discussions}
\label{sec:disscussion}

\begin{figure}
  \centerline{\includegraphics[scale=0.5]{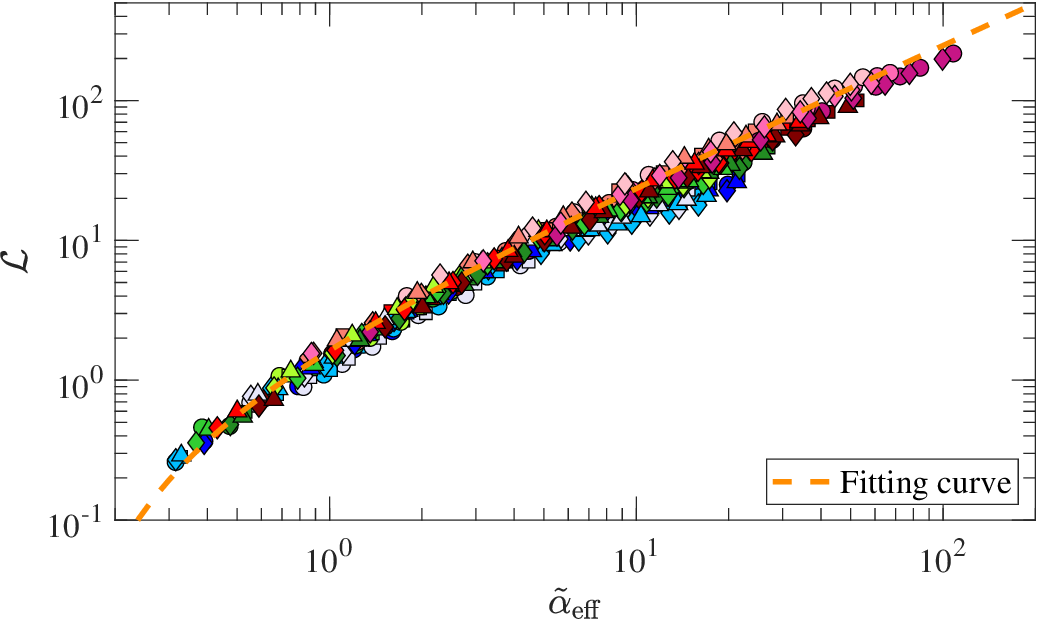}}
  \caption{Relationship between $\mathcal{L}$ and $\Tilde{\alpha}_{\rm eff}$ of granular columns with both $\phi_{\rm init} = 0.6$ and $\phi_{\rm init} = 0.8$. Markers are the same as those in Figures \ref{fig:runout0.6} and \ref{fig:runout0.8}.}
\label{fig:runoutcombined}
\end{figure}

In previous sections, we showed that by considering the extra energy input due to the inclination, the proposal of the inclined aspect ratio, $\Tilde{\alpha}_{\rm eff}$, works well in terms of describing the relative run-out behaviour of granular column collapses on inclined planes. Meanwhile, although we obtain a fair collapse of time-related variables, we have also shown the difficulties in obtaining universal descriptions for variables, such as translational and rotational kinetic energies and the maximum front velocity. Most importantly, with $\theta$, $\alpha_{\rm eff}$ and $\Tilde{\alpha}_{\rm eff}$, we are able to provide functional forms to determine the deposition height and the collapse duration of granular column collapses on inclined planes with fair accuracy. However, further discussion related to the run-out distance is still needed because the two-stage power-law relationship between $\mathcal{L}$ and $\Tilde{\alpha}_{\rm eff}$ is not completely promising and the transitional point between the two power-law relationships is unclear and equivocal. It seems that, as we continue increasing $\Tilde{\alpha}_{\rm eff}$, the $\mathcal{L} \sim \Tilde{\alpha}_{\rm eff}$ relationship will approach an asymptotic power-law solution, but with finite $\Tilde{\alpha}_{\rm eff}$, this relationship may not be a power-law function.

To resolve our concerns in the $\mathcal{L} \sim \Tilde{\alpha}_{\rm eff}$ relationship, we regard the evolution of $\mathcal{L}$ with respect to $\Tilde{\alpha}_{\rm eff}$ as a phase transition process, where granular collapses transform from quasi-static regimes to fluid-like regimes \citep{man2021deposition}. When the granular column is in a fluid-like regime, we hypothesize that the $\mathcal{L} \sim \Tilde{\alpha}_{\rm eff}$ relationship will reach a power-law asymptote. This results in a Boltzmann-like equation that
\begin{equation} \label{eq:boltzmann}
    \begin{split}
        \mathcal{L} = \kappa\Tilde{\alpha}_{\rm eff}\exp\left[ - \frac{\mathcal{E}}{(\Tilde{\alpha}_{\rm eff} - \alpha_{\rm one})^{\beta}} \right]\ ,
    \end{split}
\end{equation}
where $\kappa = 2.5$, $\mathcal{E} = 0.4$ and $\beta = 0.8$ are fitting parameters, and $\alpha_{\rm one}$ can be seen as the initial aspect ratio for granular columns with only one layer of particles so that $\alpha_{\rm one} = d_{\rm ep}/L_i$. The introduction of $\alpha_{\rm one} = d_{\rm ep}/L_i$, which may not be exact, ensures reasonably that $\mathcal{L}$ will approach 0 when $\Tilde{\alpha}_{\rm eff}$ is small enough. We combine the data in Figures \ref{fig:runout0.6} and \ref{fig:runout0.8} together and plot them in Figure \ref{fig:runoutcombined}. We plot equation \ref{eq:boltzmann} as the dashed curve in Figure \ref{fig:runoutcombined}. It shows that the simulation data agree extremely well with the proposed equation \ref{eq:boltzmann}. Further analyses are still needed to validate the Boltzmann-like equation and to find physical interpretations for parameters in Equation \ref{eq:boltzmann}.

\begin{figure}
  \centerline{\includegraphics[scale=0.5]{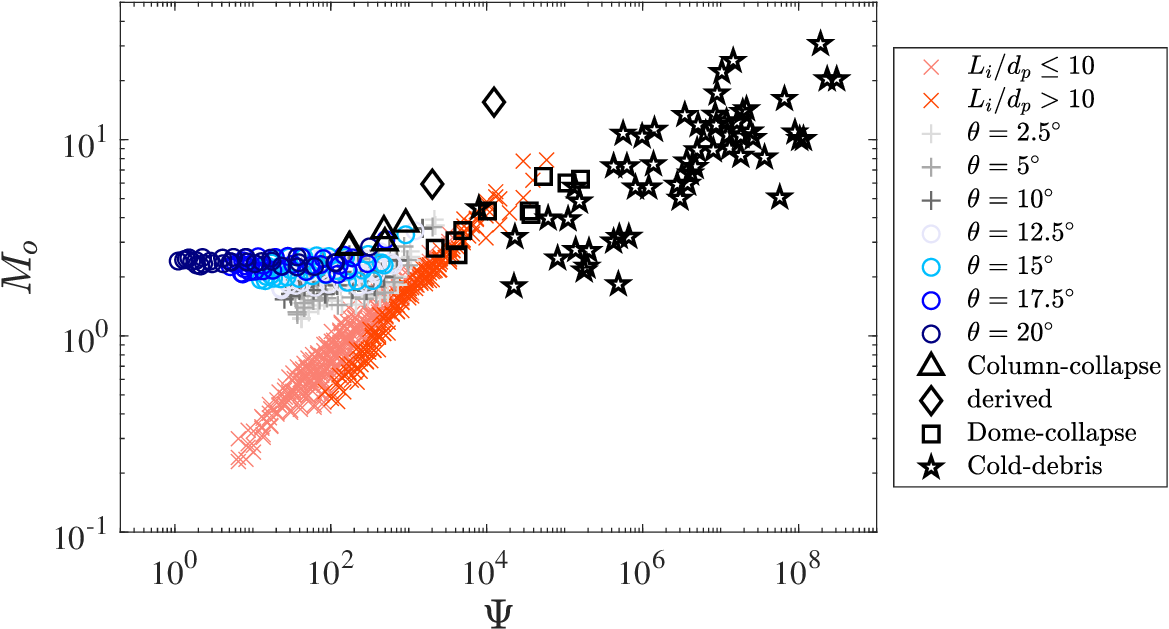}}
  \caption{Relationship between $\Psi$ and $M_o$ with comparisons of data acquired from \citet{Calder1999MobilityOP} (presented as black markers), \citet{man2021finitesize} (presented as light red $\times$ for systems with $d_p/L_i\leq 10$ and red $\times$ for systems with relative system size $d_p/L_i > 10$) and simulation results from this work ($+$ markers and blue circles for systems with different inclination angles).}
\label{fig:dataCompare}
\end{figure}

Another aspect that needs more analysis is the link between inclined granular column collapses and real granular avalanches presented in landslides or volcano-induced pyroclastic flows. \citet{Roche2002ExperimentsOD,Roche2008ExperimentalOO} investigated the correlation between dam-break granular flows and the mobility of pyroclastic flows and argued that dense and ash-rich pyroclastic flows behaved fluid-like, which is similar to some types of granular column collapses. \citet{man2021finitesize} also observed the similarity between horizontal granular column collapses in fluid-like regimes and the data of real pyroclastic flows presented in \citet{Calder1999MobilityOP}. In this work, we collect data from both \citet{Calder1999MobilityOP} and \citet{man2021finitesize}, and combine them with simulation results obtained from granular column collapses on inclined planes with various inclination angles.

In \citet{Calder1999MobilityOP}, the relationship between $M_o = L^{\prime}/H^{\prime}$ and $\Psi = \rho gV/(H^{\prime})^2$ was used to analyze qualitatively the mobility of pyroclastic flows, where $L^{\prime}$ was the collapsing distance, $H^{\prime}$ was the collapsing height that leads to different calculation methods for different types of pyroclastic flows, $\rho$ is the material density, $g$ is the gravitational acceleration and $V$ is the volume of material being transported that corresponds to $H_i L_i W_i$. For simulations presented in this work, $M_o$ is interpreted as $L_{\infty}/(H_i + \delta L\tan\theta)$ and $\Psi$ is calculated as $\rho_p g(L_i H_i W_i)/(H_i + \delta L\tan\theta)^2$. We then plot the relationship between $M_o$ and $\Psi$ along with data from \citet{Calder1999MobilityOP} and \citet{man2021finitesize} in Figure \ref{fig:dataCompare}.

In Figure \ref{fig:dataCompare}, we collect data from \citet{Calder1999MobilityOP} for different types of pyroclastic flows, such as column-collapse pyroclastic flows, derived pyroclastic flows, dome-collapse pyroclastic flows and cold-debris avalanches. For different types of pyroclastic flows, $\Psi$ varies from $10^2$ to $10^8$ due to different amounts of materials erupted. The mobility $M_o$ varies from 1 to 40. We also plot the data from \citet{man2021finitesize} as light red and red crosses to show that, as we increase the relative system size, the behaviour of horizontal granular column collapses resembles that of dome-collapse pyroclastic flows. The difference between horizontal granular column collapses and pyroclastic flows is also obvious that the slope on logarithmic coordinates of the $M_o\sim\Psi$ relationship for horizontal granular column collapses is much larger than that for pyroclastic flows. As we increase the inclination angle, the slope of the $M_o\sim\Psi$ relationship for inclined column collapses starts to decrease, which resembles the slope of natural pyroclastic flows. Results for granular systems with $\theta\in [10,\ 15]$ is similar to the behaviour of column-collapse pyroclastic flows. Additionally, we can already observe the transition from column-collapse flows to dome-collapse flows with data of granular systems with $\theta\in [10,\ 15]$. We note that, in this work, we keep the relative system size $L_i/d_p$ constant. Thus, $\Psi$ is kept at small values varying from 1 to $10^4$. We believe that, as we further increase the relative system size of granular columns, the $M_o\sim\Psi$ relationship of column collapses on inclined planes can further show similarities to other types of pyroclastic flows, which need to be addressed in future investigations.

\section{Concluding remarks}
\label{sec:conclud}

In this work, using the sphero-polyhedral discrete element simulation with Voronoi-based particles, we analyzed the behaviour of granular column collapses on inclined planes with different inclination angles varying from 2.5$^{\circ}$ to 20$^{\circ}$ to elucidate the influence of inclination angles on run-out behaviours, deposition heights, kinematics and energy transformations. Based on simulation results and their comparison with experimental data \citep{lube2011granular}, pyroclastic flow measurements \citep{Calder1999MobilityOP} and horizontal granular column collapses with different relative sizes \citep{man2021finitesize}, we draw following conclusions.

First of all, learning from \citet{man2021deposition} and based on dimensional analysis, we propose an inclined effective aspect ratio, $\Tilde{\alpha}_{\rm eff}$, to address both the extra potential energy a granular column can utilize for a longer run-out distance and the reduction of frictional effect due to the inclination. With $\Tilde{\alpha}_{\rm eff}$, we gain great advantages in describing the run-out distance for both experimental results reported by \citet{lube2011granular} (Figure \ref{fig:expres}) and the simulation data in this work (Figures \ref{fig:runout0.6} and \ref{fig:runout0.8}). In Section \ref{sec:disscussion}, we further link the relationship between $\mathcal{L}$ and $\Tilde{\alpha}_{\rm eff}$ to a Boltzmann-like equation to show that, as we increase $\Tilde{\alpha}_{\rm eff}$, the $\mathcal{L}\sim\Tilde{\alpha}_{\rm eff}$ relationship approaches a power-law asymptote.

We also show that the dimensionless collapse duration $T_f/\sqrt{L_i/g}$ is strongly correlated with $\Tilde{\alpha}_{\rm eff}$, and $T_f/\sqrt{L_i/g}$ exhibits power-law relationships with $\Tilde{\alpha}_{\rm eff}$ given by $T_f/\sqrt{L_i/g} = \mathcal{A}_{\rm tf}\Tilde{\alpha}_{\rm eff}^{0.4}$, but $\mathcal{A}_{\rm tf}$ is still a function of the inclination angle $\theta$. This indicates that the collapse duration and the inclination angle have a complex relationship and that $\Tilde{\alpha}_{\rm eff}$ alone is not able to fully determine the collapse duration. However, the times for a column collapse to reach its maximum translational kinetic energy, its maximum rotational kinetic energy and its maximum front velocity ($T_{\rm kt,max}$, $T_{\rm ka,max}$ and $T_{\rm ufr,max}$) can all be written as power-law functions of the inclined effective aspect ratio $\Tilde{\alpha}_{\rm eff}$, which all indicate the useful side of $\Tilde{\alpha}_{\rm eff}$ in determining the run-out behaviour and time-related variables. Similarly, the final deposition height can be determined based on the relationship between $0.5H_{\infty}L_{\infty}/(H_i L_i)$ and $\alpha_{\rm eff}$, where $0.5H_{\infty}L_{\infty}/(\mathcal{F}(\theta)H_i L_i)$ can be written as a function of $\alpha_{\rm eff}$. We can clearly see that a separation of variables can be performed while calculating $0.5H_{\infty}L_{\infty}/(H_i L_i)$ that the influence of the inclination angle and the frictional interaction are independent from each other. 

Meanwhile, both the maximum kinetic energy and the maximum front velocity seem to be insensitive to the inclination angles. The effective aspect ratio $\alpha_{\rm eff}$ alone can give a reasonable prediction of the maximum translational and rotational kinetic energies and the maximum front velocity. We conclude that the change of the inclination angle, which transforms more potential energy into kinetic energies, mainly results in a longer duration for energy transformation instead of promoting a larger maximum kinetic energy and a larger front velocity. This implies that, for natural granular avalanches on slopes with different inclinations, it may be more important to consider the resulting flowing duration and the increase of flooded area than to accurately calculate the damage it can cause to a single structure (that is more or less governed by the maximum front velocity of the flow). This investigation covers the broad topic of granular columns collapses on inclined planes with the proposal of utilizing both $\Tilde{\alpha}_{\rm eff}$, $\alpha_{\rm eff}$, and $\theta$ to predict the propagation length and propagation duration of granular collapses, which is of vital importance to better understand the fundamental physics behind some natural geophysical flows. This work focuses on discrete element simulations with Voronoi-based grains but simple boundary conditions, which differs from some granular-like flows in natural and engineering systems. Thorough investigations are still needed to explore more complicated situations and to elucidate the impact of granular collapses with different boundary conditions. We plane to take these up in our future studies.

\begin{acknowledgements}
\textbf{Acknowledgements}- The authors acknowledge the financial support from the National Natural Science Foundation of China with project number 12202367 and 12172305, and thank Westlake University and the Westlake High-performance Computing Center for computational and experimental sources and corresponding assistance. T.M. would like to acknowledge the helpful discussions with Prof. K. M. Hill from the University of Minnesota.
\\
\textbf{Declaration of Interests}- The authors report no conflict of interest.
\end{acknowledgements}

\bibliographystyle{jfm}
\bibliography{collapseInclined}

\end{document}